\documentclass{jkas}

\def\year{2021} 
\def\month{xx} 
\def\volume{00} 
\def\issue{0} 
\def\beginpage{1} 
\def\endpage{99} 
\def\received{August xx, 2021} 
\def\accepted{xx xx, 2021} 
\date{Received \received; accepted \accepted}

\usepackage{makecell}
\usepackage{multirow}
\usepackage{ulem}
\usepackage{xcolor}
\usepackage{float}
\newcommand{\kms}{km s$^{-1}$}
\newcommand{\dego}{$^\circ$}
\newcommand{\msun}{M$_\odot$}

\title{
ALMA observations of W Hya: impact of missing baselines
}

\author[]{Do Thi Hoai}
\author[]{Pham Tuan Anh}
\author[]{Pham Tuyet Nhung}
\author[]{Pierre Darriulat}
\author[]{Pham Ngoc Diep}
\author[]{Nguyen Bich Ngoc}
\author[]{Tran Thi Thai}

\affil{Department of Astrophysics, Vietnam National Space Center, Vietnam Academy of Science and Technology, 18, Hoang Quoc Viet, Nghia Do, Cau Giay, Ha Noi, Vietnam. \email{dthoai@vnsc.org.vn, ptanh@vnsc.org.vn}}

\def\corrauthor{
D.T. Hoai and P. Tuan-Anh
}

\def\runningauthor{
D.T Hoai et al.
}

\def\runningtitle{
W Hya: impact of missing baselines
}

\def\keywords{
stars: AGB and post-AGB -- circumstellar matter -- stars: individual: W Hya -- radio lines: stars -- techniques: image processing, interferometric 
}

\def\abstracttext{
The lack of short baselines, referred to as short-spacing problem (SSP), is a well-known limitation of the performance of radio interferometers, causing a reduction of the detected flux from large scale source structures. The very large number of antennas operated in the Atacama Large Millimeter/sub-millimeter Array (ALMA) generates situations for which the impact of the SSP takes a complex form, not simply measurable by a single number, such as the maximal recoverable scale. In particular extended antenna configurations, complemented by a small group of closeby antennas at the centre of the array, may result in a double-humped baseline distribution with a significant gap between the two groups. In such cases one should adopt as effective maximal recoverable scale that associated with the extended array and use only the central array to recover missing flux, as one would do with single dish or ACA (Atacama Compact Array) observations. The impact of the missing baselines can be very important and may easily be underestimated, or even overlooked. The present study uses ALMA archival data of the $^{29}$SiO(8-7) line emission of AGB star W Hya as an illustration. A critical discussion of the reliability of the observations away from the star is presented together with comments of a broader scope. Properties of the circumstellar envelope of W Hya within $\sim$15 au from the star, many of which are not mentioned in the published literature, are briefly described and compared with R Dor, an AGB star having properties very similar to W Hya.
}

\begin{document}
\jkashead 


\section{Introduction}{\label{sec1}}

In the recent years, the operation of the Atacama Large Millimeter/submillimeter Array (ALMA) has opened a new era of radio interferometry with an unprecedented number of antennas, offering a broad choice of possible configurations. In particular, the availability of long baselines has often favoured high angular resolution configurations at the detriment of short baselines, resulting in significant losses of detected flux from large source structures. This so-called short-spacing problem (SSP) is well known and has been the subject of numerous studies \citep{Faridani2018, Braun1985}. In particular, incomplete coverage of the $uv$ plane is known to produce artefacts known as ``ghosts'' or ``invisible distributions'' \citep[][and references therein]{Chandra2012}. The SSP is typically coped with by merging the array data with single dish observations, or, in the case of ALMA, with data collected using the compact array (ACA). However, the high diversity of possible antenna configurations producing short baselines may result in a complex morphology of the map of the array acceptance, the precise knowledge of which is mandatory for a reliable interpretation of the data. In particular, it may not be reducible to a single number, the so-called maximal recoverable scale. In principle, ALMA users are properly warned of this fact and are encouraged to produce simulations of the imaging process for the specific antenna configuration being used (see for example Chapter 7 of the ALMA technical handbook\footnote{https://almascience.nao.ac.jp/documents-and-tools/cycle7/alma-technical-handbook}). In practice, however, it is often easy to underestimate the importance of this measure and the danger to overlook significant imaging distortions is real. We illustrate the argument using archival ALMA observations of the emission of the $^{29}$SiO($\nu$=0, $J$=8-7) line by the circumstellar envelope (CSE) of W Hya, an oxygen-rich AGB star at a distance of only $\sim$104$^{+14}_{-11}$ pc to the Sun \citep{VanLeeuwen2007}.  It is a semi-regular variable with a period of $\sim$361 days \citep{Samus2017}, often quoted as a Mira \citep{Lebzelter2005} belonging to spectral class M7.5e-M9ep. Its mass-loss rate is $\sim$10$^{-7}$ \msun yr$^{-1}$ \citep{Maercker2008, Khouri2014a}. Its main sequence mass was between 1 and 1.5 \msun\ \citep{Khouri2014b, Danilovich2017}.

\begin{figure*}
  \centering
  \includegraphics[height=4.4cm,trim=0.2cm .5cm 0.2cm 1.7cm,clip]{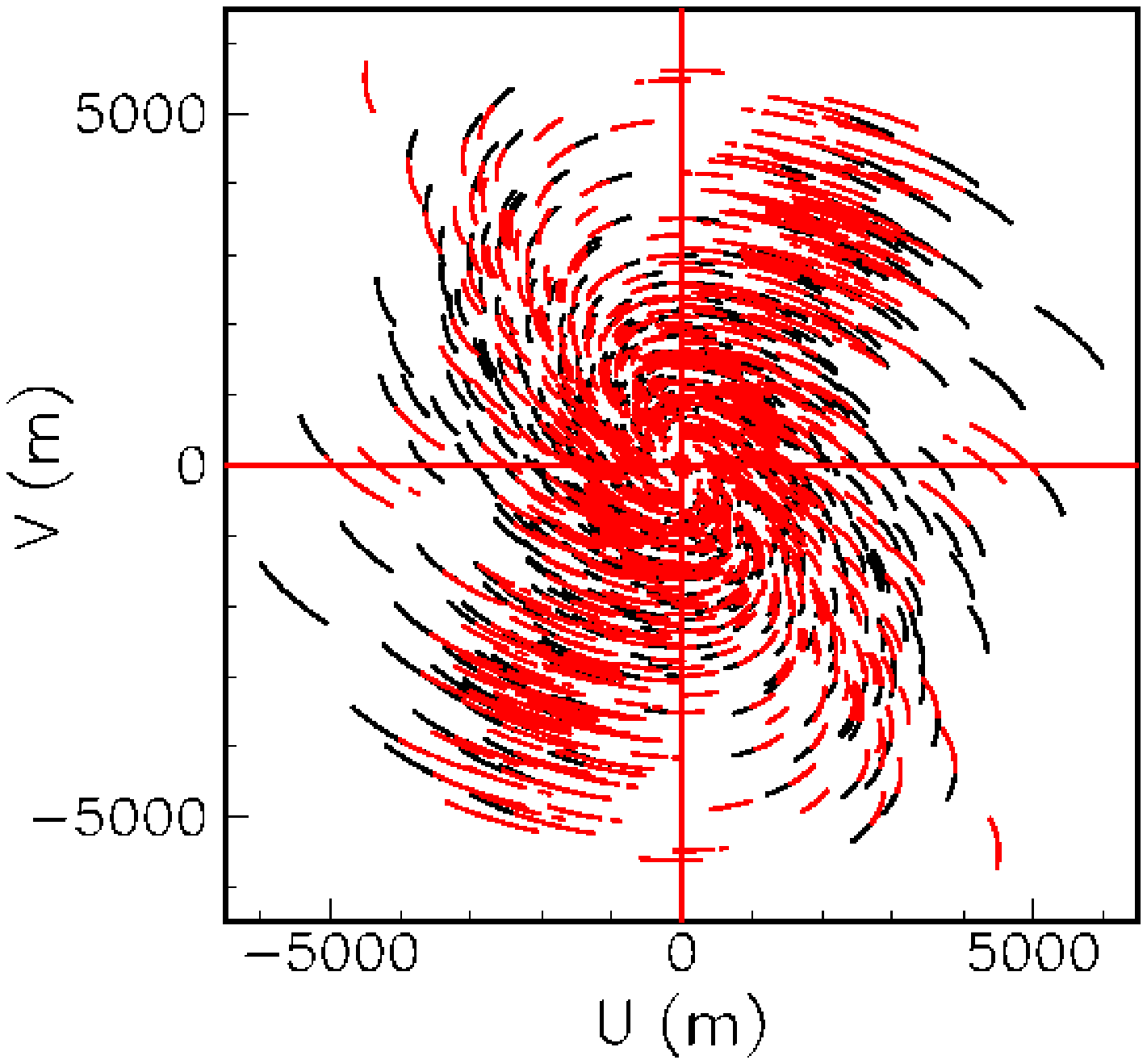}
  \includegraphics[height=4.4cm,trim=0.2cm .5cm 0.2cm 1.7cm,clip]{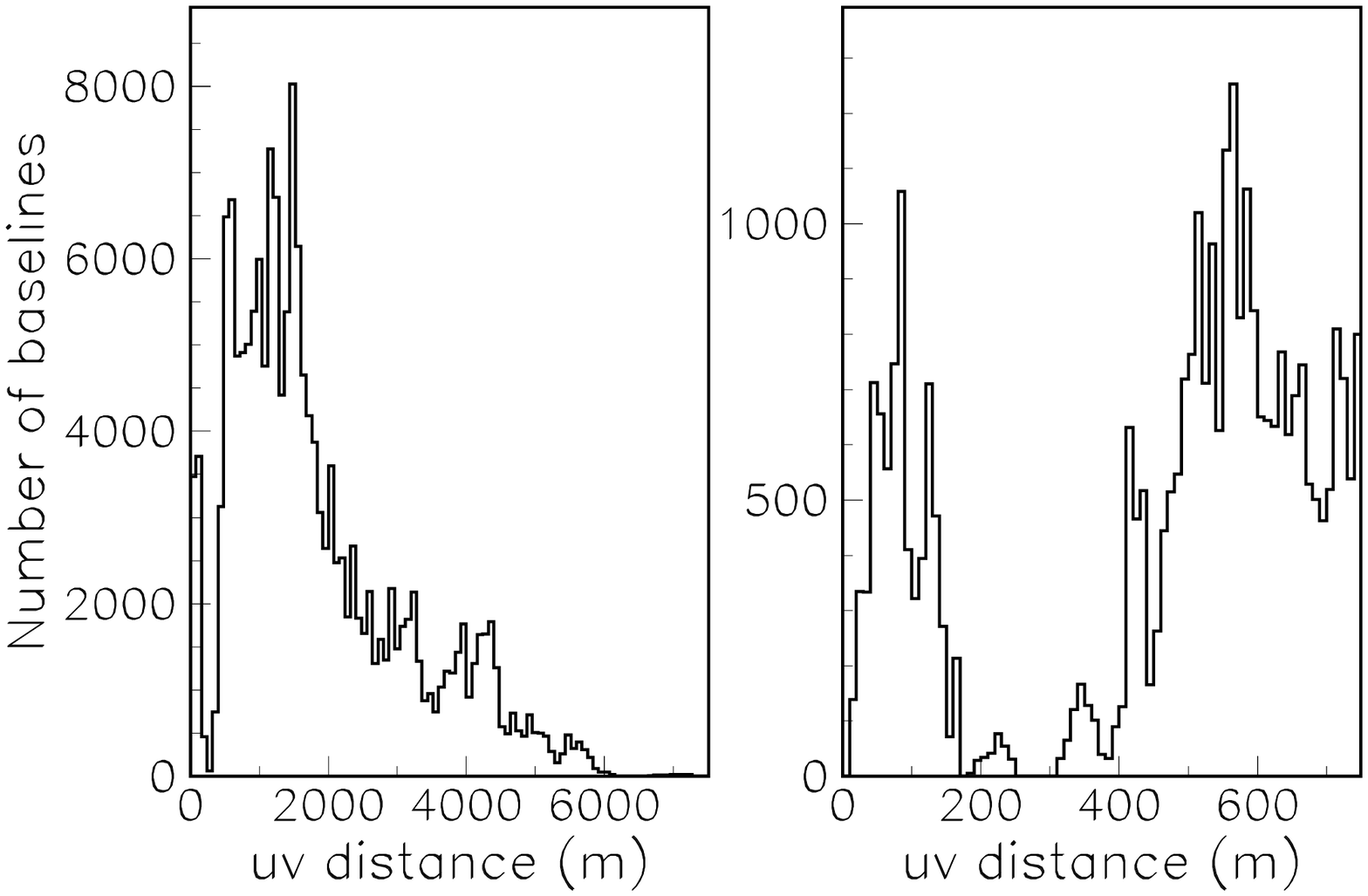}
  \includegraphics[height=4.4cm,trim=-.5cm .5cm 0.2cm 1.5cm,clip]{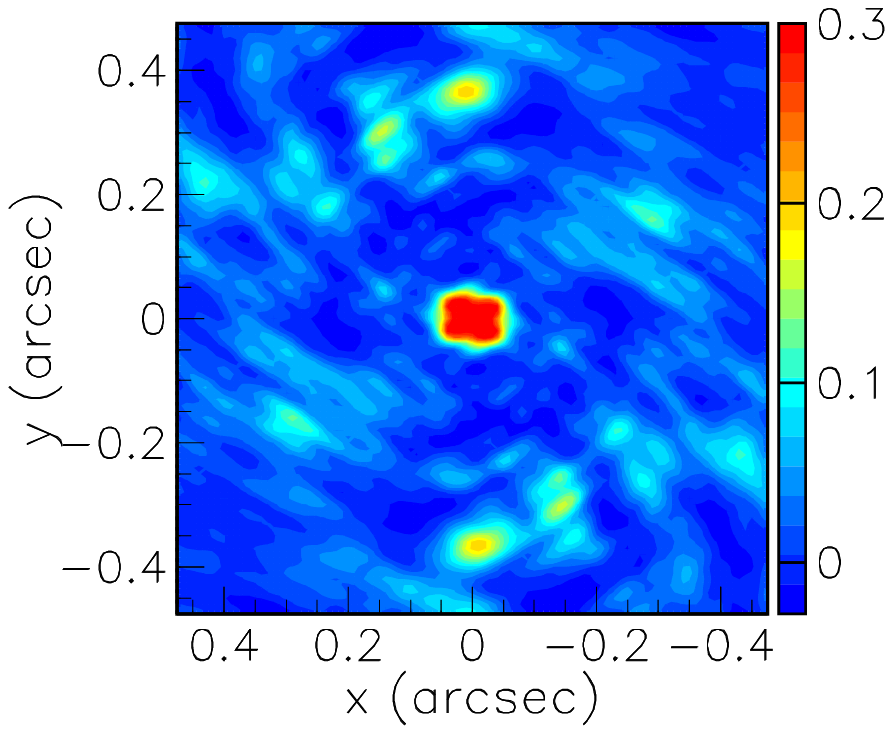}\\
  \includegraphics[height=4.4cm,trim=0.2cm .5cm 0.2cm 1.5cm,clip]{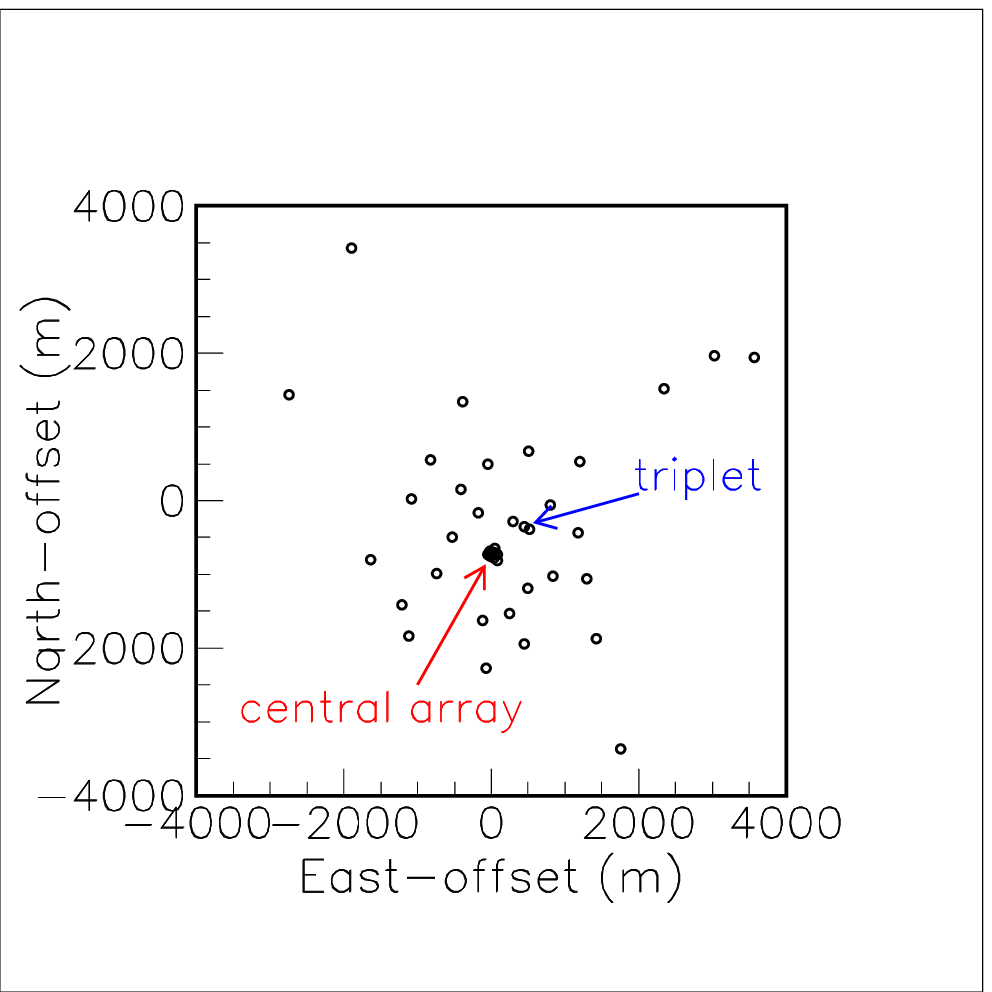}
  \includegraphics[height=4.4cm,trim=0.2cm .5cm 0.2cm 1.5cm,clip]{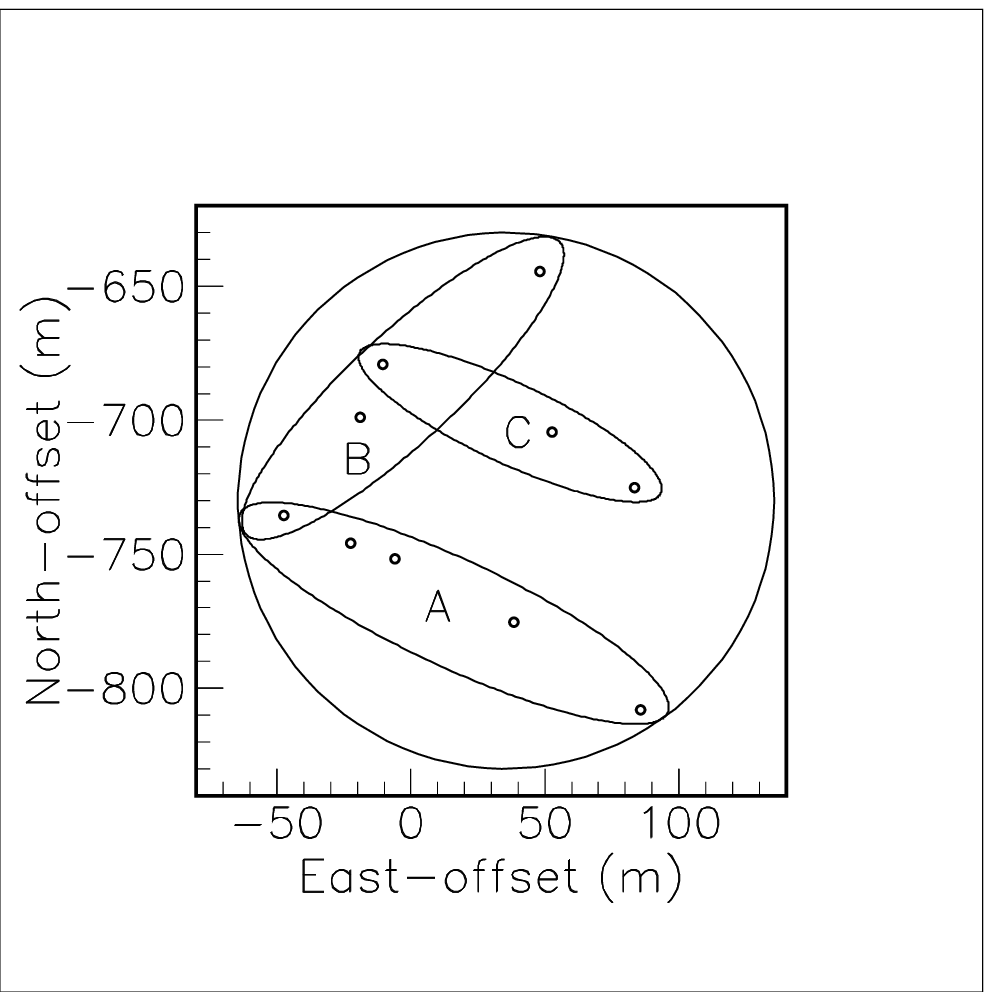}
  \includegraphics[height=4.4cm,trim=0.2cm .5cm 0.2cm 1.5cm,clip]{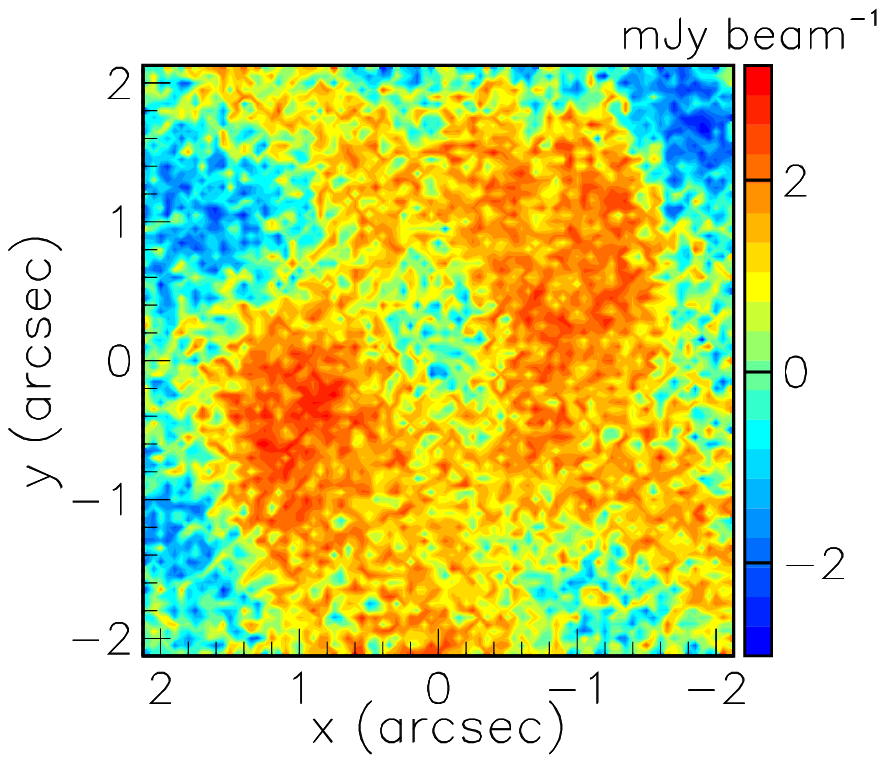}
  \caption{Upper panels: $uv$ coverage for blocks 1+2. In the left panel, block1 is black and block 2 is red. Baseline distributions are shown in the central panels, the rightmost being a zoom on smaller baselines with a bin size of 10 m instead of 80 m. Right: Dirty beam (natural weighting); the colour map is saturated at 0.3 (rather than 1) to show the side lobe of the dirty beam. Lower panels: antenna configuration for blocks 1+2 (42 antennas); the central panel is a zoom on the central array; the red arrow (left) points to the central array, the blue arrow points to a triplet of antennas of the extended array producing short baselines; the circle (centre) has a radius of 100 m; the ellipses define three antenna alignments: A, B and C that are referred to in Figure \ref{fig7}. Right: map of residuals (natural weighting) of a frequency channel close to systemic velocity; the maximum of the colour scale is $\sim$3 mJy beam$^{-1}$, 1.4 times the 1-$\sigma$ level of the noise (2.1 mJy beam$^{-1}$).}
 \label{fig1}
\end{figure*}

\section{Observations, data reduction and imaging}{\label{sec2}}

The present work uses archival observations of W Hya from project ADS/JAO.ALMA\#2015.1.01446.S (PI: A. Takigawa), which were carried out for a total of $\sim$2 hours on source over three separate blocks between 30 November and 5 December 2015 with ALMA in Cycle 3. In the present work we use mostly blocks 1 and 2, the data of block 3 being of lesser quality, but we have checked that all the arguments made in the article are independent of the blocks being used, 1+2+3 or 1+2 or 2 alone, each block displaying very similar antenna configurations. The antennas, respectively 33, 41 and 31 in number, were configured in such a way that the baseline lengths were distributed in two groups, one covering between $\sim$15 m and $\sim$200 m, the other between $\sim$400 m and $\sim$8 km; the former group included 10 antennas in a circle of 100 m radius (Figure \ref{fig1}). The emission of the $^{29}$SiO (8-7) line, with a frequency of 342.9808 GHz (wavelength $\lambda$$=$0.875 mm), was covered with a frequency resolution of $\sim$977 kHz channel$^{-1}$, corresponding to a channel spacing of 0.854 \kms. The data have been reduced using standard scripts without continuum subtraction, with particular attention to the scale over which the flux is reliably recoverable. 
Both imaging codes CASA and GILDAS have been used\footnote{https://casa.nrao.edu and https://imager.oasu.u-bordeaux.fr/}.  We have checked the consistency between the results obtained with the two codes and with different parameters of the deconvolution algorithms, including weights, masks, number of iterations, etc. In all cases, a proper behaviour was observed and the arguments developed in the article were found to be valid independently from these. As an illustration, we show in Figure \ref{fig1} the dirty beam and a map of residuals obtained with GILDAS using natural weighting. Imaging produces a beam of 63$\times$52 mas$^2$ with PA=79\dego\ with natural weighting and a beam of 52$\times$38 mas$^2$ with PA=99\dego\ with robust weighting (threshold of 1).

\begin{figure*}
  \centering
  \includegraphics[height=5cm,trim=0.0cm .5cm 1.5cm 1.cm,clip]{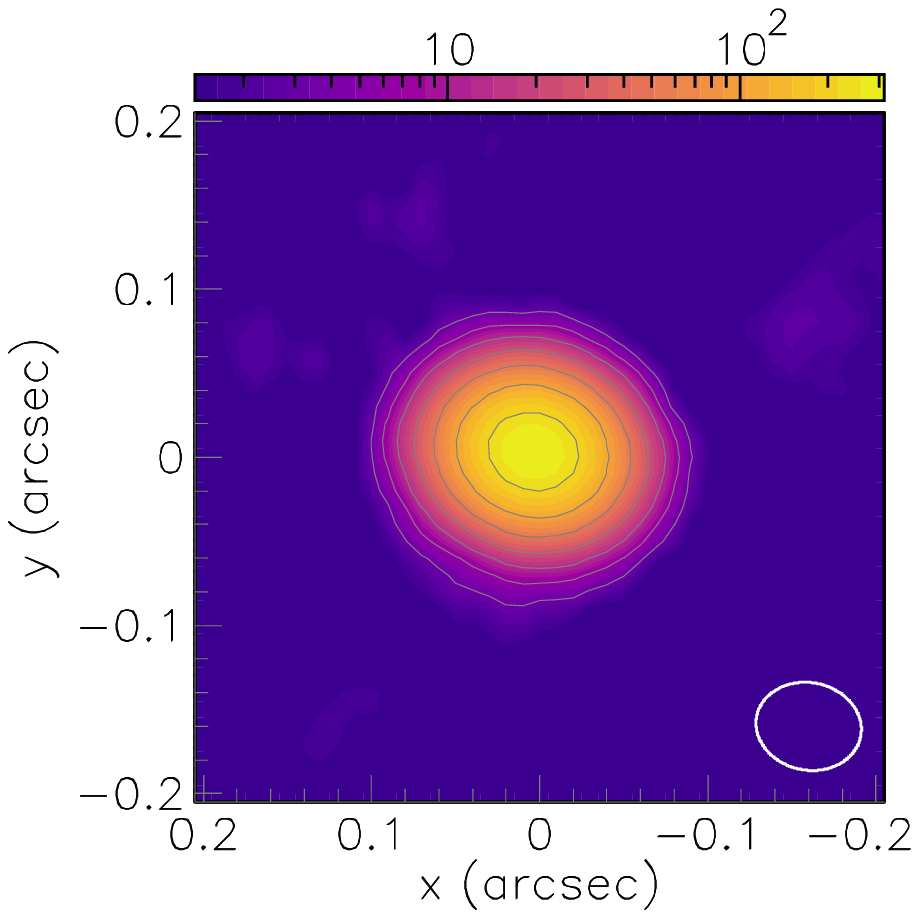}
  \includegraphics[height=5cm,trim=1.5cm .5cm 1.5cm 1.cm,clip]{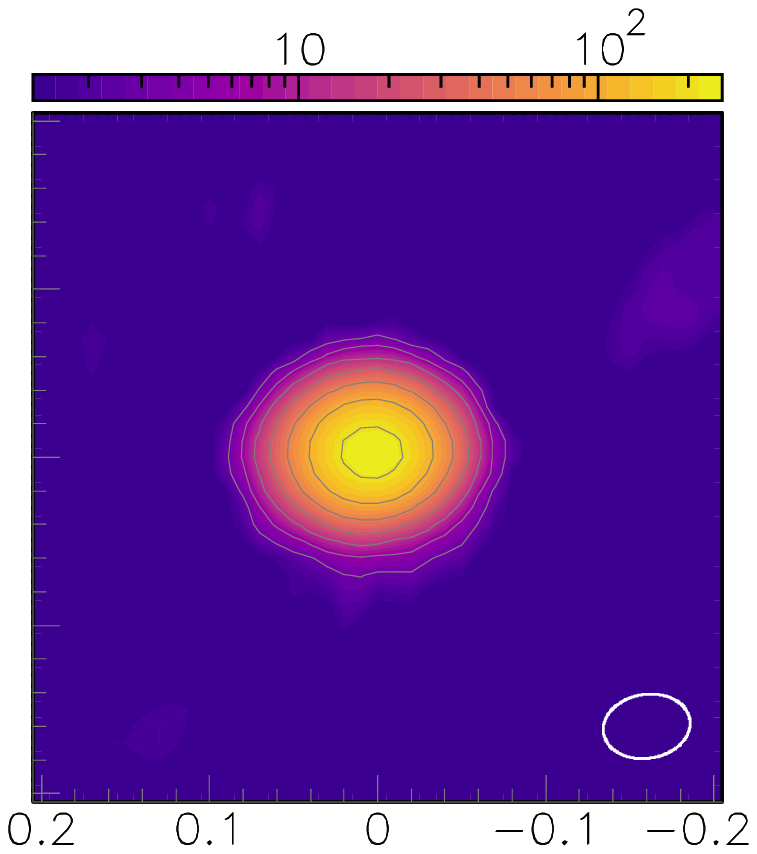}
  \caption{Maps of the continuum brightness obtained with natural (left) and robust (right) weighting. The colour scale is in units of mJy\,beam$^{-1}$. The beam is shown in the lower right corner of each panel.}
 \label{fig2}
\end{figure*}

\begin{figure*}
  \centering
  \includegraphics[width=5cm,trim=0.0cm .5cm 1.cm 1.5cm,clip]{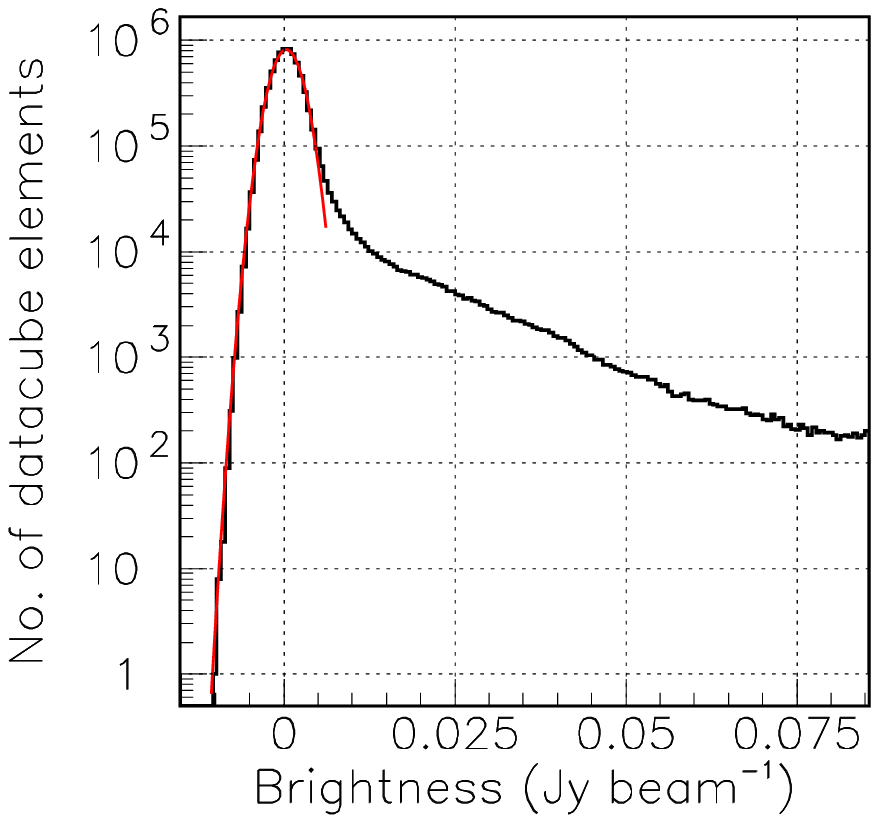}
  \includegraphics[width=5cm,trim=0.0cm .5cm 1.cm 1.5cm,clip]{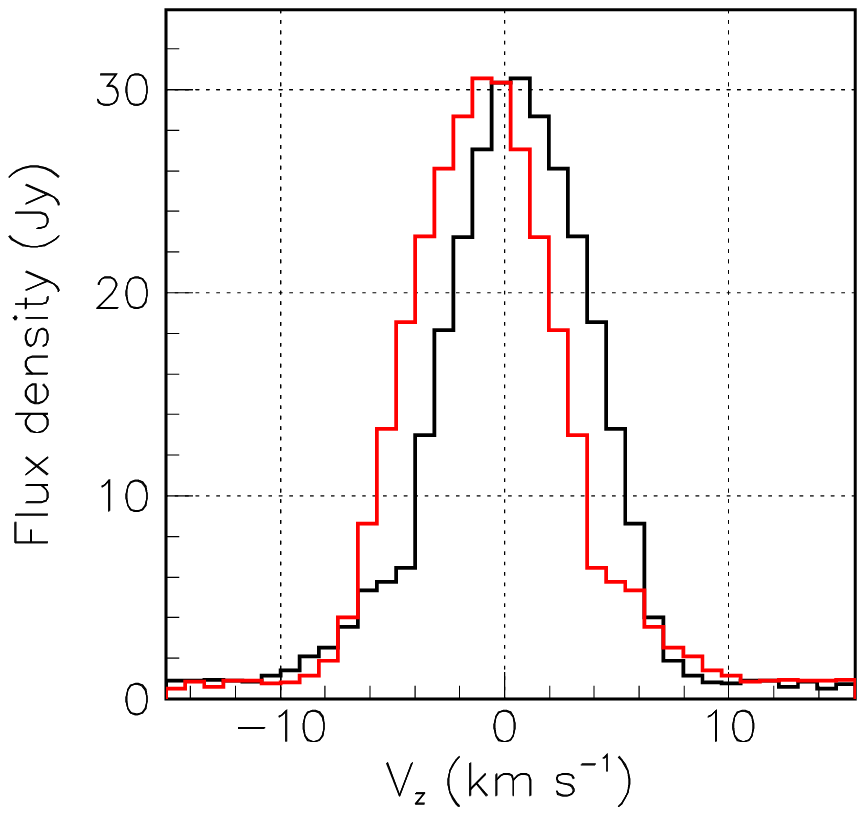}
  \includegraphics[width=5cm,trim=0.0cm .5cm 1.cm 1.5cm,clip]{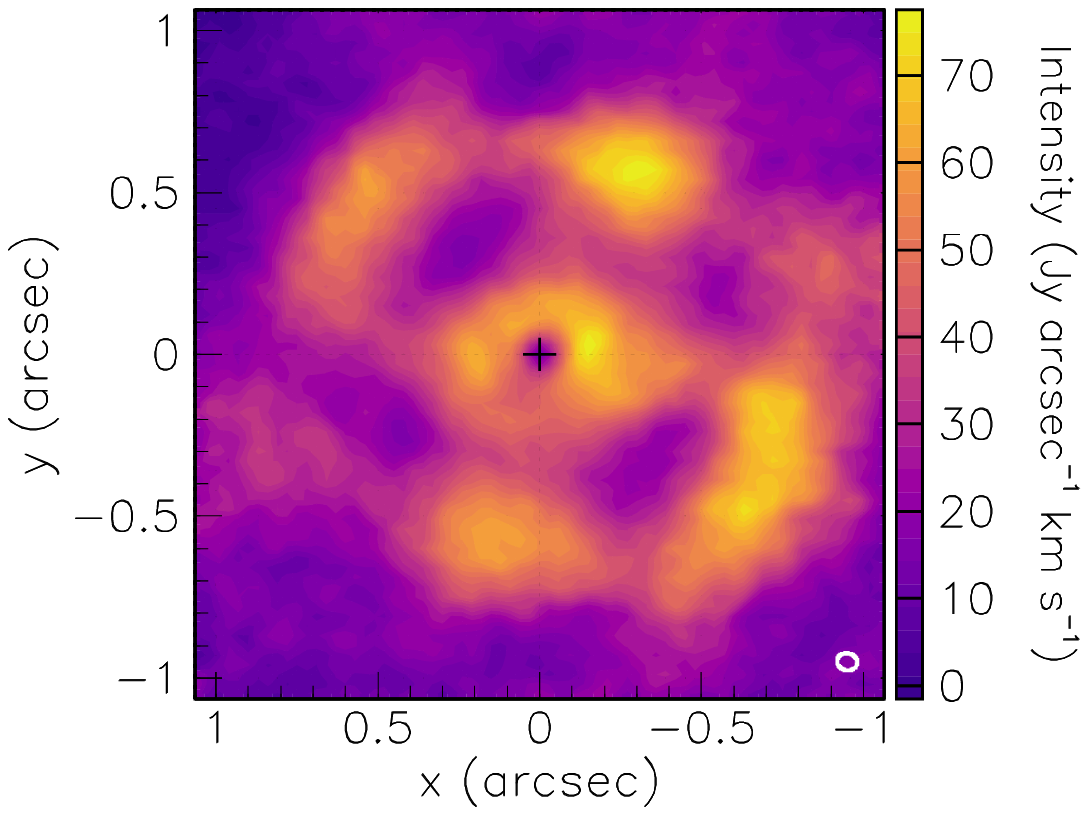}
  \caption{ Left: brightness distribution in a $R$$<$2 arcsec circle (natural weighting); the red line shows the result of a Gaussian fit. Middle: Doppler velocity spectrum integrated over $R$$<$1 arcsec; the asymmetry, illustrated by the mirror spectrum shown in red, is the result of the absorption centred at the terminal wind velocity of $\sim-$5.5 \kms\ (see Figure \ref{fig9}). Right: map of the intensity integrated over $|V_z|$$<$8 \kms\ and multiplied by $R$. The beam is shown in the lower right corner.}
 \label{fig3}
\end{figure*}

Maps of the continuum emission are shown in Figure \ref{fig2}. After beam deconvolution, Gaussian fits to the stellar disc give a diameter of 35$\pm$2 mas FWHM, in good agreement with earlier evaluations \citep{Vlemmings2019}. We use coordinates centred on the continuum emission, $x$ pointing east, $y$ pointing north and $z$ pointing away from Earth. The projected distance to the star is calculated as $R=\sqrt{x^2+y^2}$. Position angles, $\omega$, are measured counter-clockwise from north. The Doppler velocity $V_z$ spectrum is centred on a systemic velocity of 40.4 \kms.

\begin{figure*}
  \centering
  \includegraphics[width=6.cm,trim=0.0cm 1cm 0.cm 1.cm,clip]{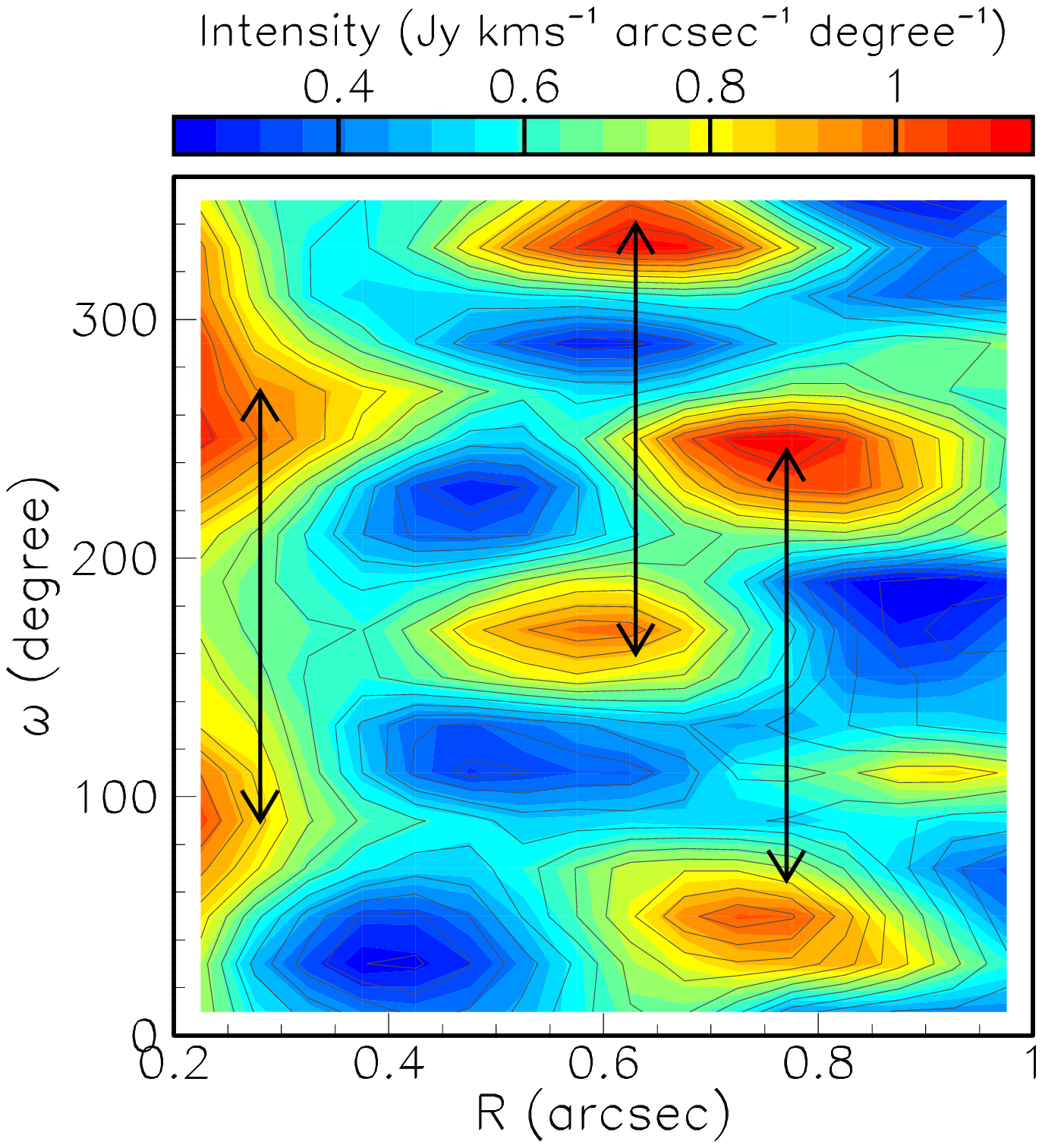}
  \includegraphics[width=6.cm,trim=0.0cm 1.cm 0.cm 1.cm,clip]{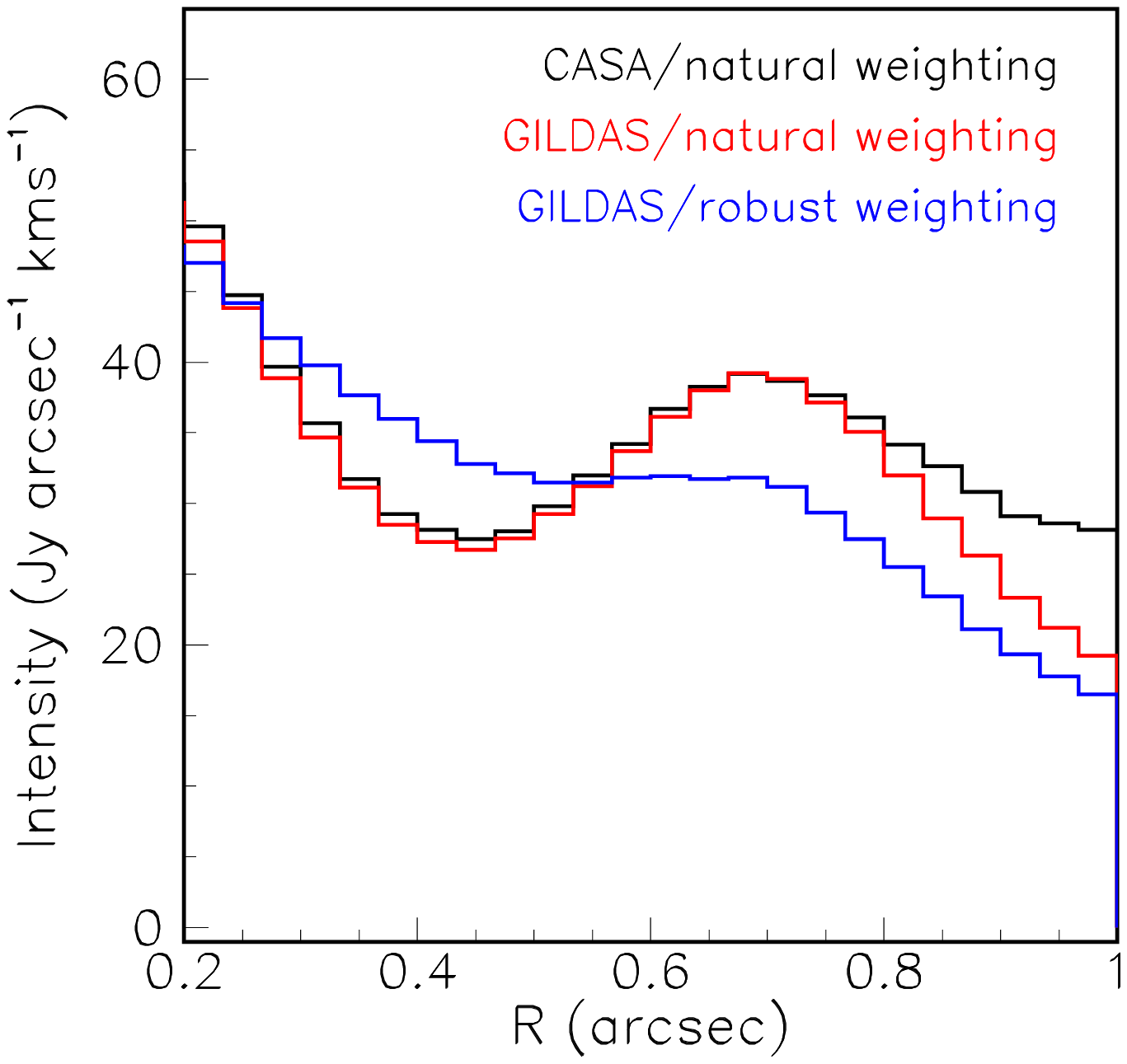}
  \caption{Comparison between the clean maps (integrated over $|V_z|$$<$5 \kms) obtained using different deconvolution algorithms. Left: clean map in the $\omega$ vs $R$ plane reconstructed using CASA (colour) and using GILDAS (contours). The arrows show a spacing of 180\dego\ in position angle. Right: $R$ distributions integrated over position angle . }
 \label{fig4}
\end{figure*}

Figure \ref{fig3} displays the brightness distribution of the $^{29}$SiO(8-7) emission in a $R$$<$2 arcsec circle centred at the origin; a Gaussian fit to the noise gives a $\sigma$ of 2.1 mJy\,beam$^{-1}$. Also shown are the Doppler velocity spectrum and the intensity map multiplied by $R$, the latter displaying a complex pattern that is further illustrated in Figure \ref{fig4}, using polar rather than Cartesian sky coordinates. The observed pattern shows back-to-back outflows and a depression around $R$$\sim$0.45 arcsec, deeper when using natural weighting than when using robust weighting.  Overall, it suggests successive emission in a plane close to the plane of the sky, at intervals of typically 0.25 arcsec, meaning some 30 years for a projected expansion velocity of 4 km s$^{-1}$ (see Section \ref{sec4}), of three successive pairs of back-to back outflows differently oriented on the plane. Such a pattern has never been observed in earlier studies of AGB stars; we show in the next section that it is an artefact of the particular antenna configuration.

\section{Impact of the two-humped baseline distribution}{\label{sec3}}
\subsection{General considerations}
Before addressing the specific case of the present W Hya observations, we briefly recall some basic features of radio interferometry. The complex visibility distribution in the $uv$ plane, $V(u,v)$, is defined as the Fourier transform of the source brightness distribution in the sky plane, $I(x,y)$:

\begin{eqnarray}
\label{eq:uv}
V(u,v)&=&\int{I(x,y)e^{-2i\pi(ux+vy)} \mathrm{d}x \mathrm{d}y} \nonumber\\
I(x,y)&=& \int{V(u,v)e^{2i\pi(ux+vy)} \mathrm{d}u \mathrm{d}v} \nonumber
\nonumber
\end{eqnarray}

A specific antenna configuration, operated for a given period of Earth rotation, samples only part of the complex visibility distribution in the $uv$ plane, defined by a sampling function $S(u,v)$, a sum of two-dimensional delta-functions. The Fourier transform of the measured visibilities, $I_D(x,y)$, called ``dirty map", is the convolution of the source brightness distribution by the Fourier transform of the sampling function, $B(x,y)$, called Point Spread Function or ``dirty beam": 
\begin{eqnarray}
\label{eq:uv2}
I_D(x,y)&=&\int{S(u,v)V(u,v)e^{2i\pi(ux+vy)} \mathrm{d}u \mathrm{d}v}\nonumber\\
&=&\int{B(x-x',y-y')I(x',y') \mathrm{d}x'\mathrm{d}y' }\nonumber
\nonumber
\end{eqnarray}

Accordingly, extended distributions of baseline lengths produce high angular resolutions, the higher the longer the longest baselines. Conversely, the shortest baselines define the maximal recoverable scale beyond which flux fails to be detected. The definition of the shortest and longest baselines is somewhat arbitrary, depending on how progressively the baseline length distribution rises and falls; moreover, it assumes a reasonably smooth baseline distribution: the presence of peaks and/or gaps impacts the dirty map in a complex way that cannot be summarised by single numbers such as angular resolution and maximal recoverable scale.

Quantitatively, a source brightness distribution $I'(x,y)=k^2I(kx,ky)$, obtained by shrinking $I(x,y)$ by a factor $k$ in the plane of the sky, produces a visibility distribution in the $uv$ plane expanded by the same factor $k$:
\begin{eqnarray}
\label{eq:uv3}
V'(u,v)&=&k^2\int{I(kx,ky)e^{-2i\pi(ux+vy)} \mathrm{d}x \mathrm{d}y} \nonumber\\
&=&\int{I(x',y')e^{-2i\pi[(u/k)x'+(v/k)y']} \mathrm{d}x' \mathrm{d}y'} \nonumber\\
&=&V(u/k,v/k) \nonumber
\end{eqnarray}

Similarly, an antenna pattern obtained by shrinking the pattern described by the sampling function $S(u,v)$, itself described by a sampling function $S'(u,v)=k^2S(ku,kv)$, produces a dirty map:
\begin{eqnarray}
\label{eq:uv4}
&&I'_D(x,y)=k^2 \int{S(ku,kv)V(u,v)e^{2i\pi(ux+vy)} \mathrm{d}u \mathrm{d}v}\nonumber\\
&=&\int{S(u',v')V(u'/k,v'/k)e^{2i\pi[u'(x/k)+v'(y/k)]} \mathrm{d}u' \mathrm{d}v'}\nonumber\\
&=&\int{S(u',v')V'(u',v')e^{2i\pi[u'(x/k)+v'(y/k)]} \mathrm{d}u' \mathrm{d}v'}\nonumber
\end{eqnarray}

Namely, the dirty map produced by an antenna pattern shrunk by a factor $k$ is the same as the dirty map obtained by expanding by a factor $k$ the dirty map of the source shrunk by the same factor $k$.
Whatever visibility is not sampled by the sampling function does not contribute to the dirty map. Producing a clean map implies therefore assigning values to the complex visibility in the regions of the $uv$ plane that are not sampled by $S(u,v)$. Arbitrary visibilities in such regions have no effect on the dirty map 
but cause the Fourier transform of the visibility distribution to arbitrarily deviate from the source brightness distribution.

In the present case of a two-humped baseline distribution, we can approach the study of the measurement process from two different points of view. A first approach is to consider the extended array as the basic array and to study the contribution obtained by adding the central array. This is reminiscent of what is commonly done when merging interferometer data with single dish data, or, in the case of ALMA, main 12 m array data with ACA data. A second approach is to consider as basic array an array that would include the complete array, central+extended, as well as additional antennas filling the baseline gap between the two; one would then study the impact of removing these additional antennas and would speak about the effect of ``missing baselines". In the present work, we may take one or the other approach: our ambition is not to produce an exhaustive treatment of the problem but simply to illustrate with a specific example the danger of underestimating its importance and of unduly using concepts such as maximal recoverable scale or angular resolution, which are insufficient to properly describe the reality of the measurement process.

\subsection{W Hya observations: radial distributions}
\begin{figure*}
  \centering
  \includegraphics[width=6.cm,trim=0.0cm 1.cm 0.cm 1.cm,clip]{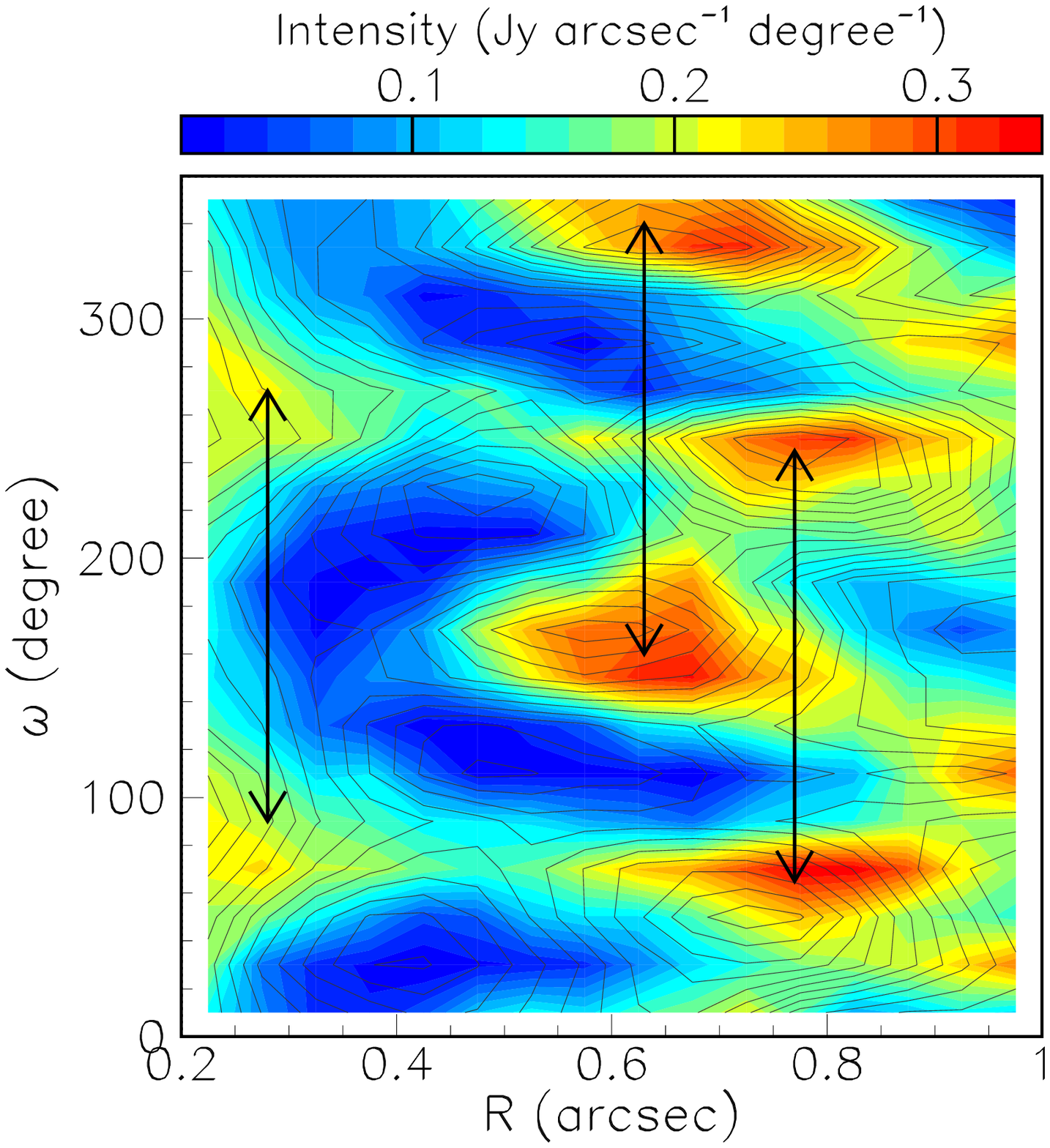}
  \includegraphics[width=6.cm,trim=0.0cm 1.cm 0.cm 1.cm,clip]{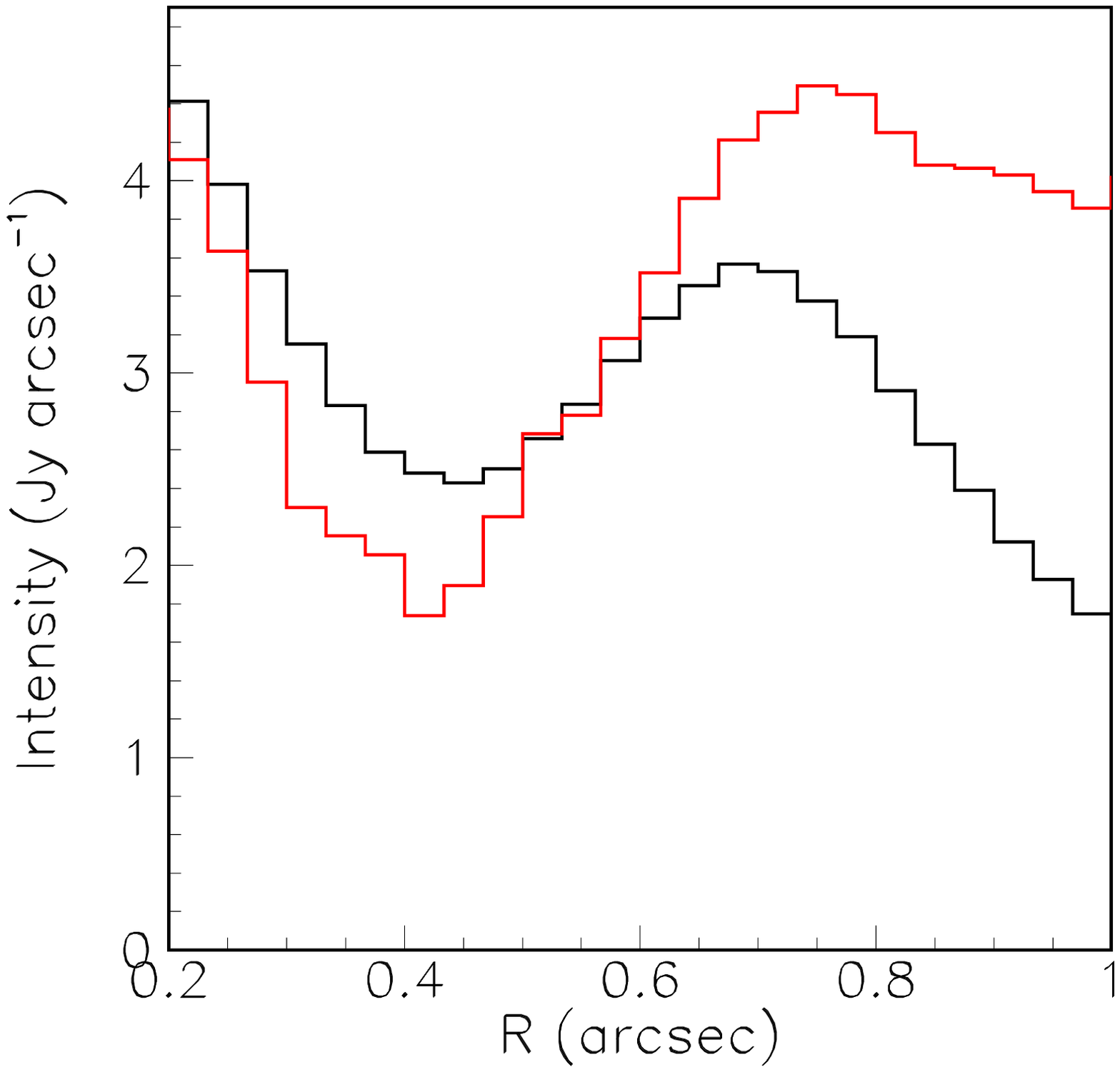}
  \caption{Imaging the isotropic wind model of emission with the antenna configuration of data sets 1 and 2. Left: clean map ($\omega$ vs $R$) of the model intensity, contours are from the data. Right: $R$ distributions (red is for the model, black is for the data). The data are averaged over $|V_z|$$<$5 \kms.}
 \label{fig5}
\end{figure*}

In order to evaluate the impact of the two-humped baseline distribution on the results presented in Section \ref{sec2} and to estimate the projected distance from the star up to which they can be considered reliable, we simulate the response of the array to the emission of an optically thin isotropic wind of constant radial velocity, with an intensity decreasing in inverse proportion to the projected distance to the star, $R$. To avoid the singularity of such a wind at $R$=0, we use a uniform brightness of 72 Jy arcsec$^{-2}$ within the disc $R$$<$0.1 arcsec. As the observed pattern is seen to evolve smoothly from frequency channel to frequency channel, it is sufficient at this stage to ignore the dependence on Doppler velocity. The clean image obtained from the visibilities produced by the actual antenna configuration is shown in Figure \ref{fig5} and displays a pattern strikingly similar to that observed for the real data: a same depression around $R$$\sim$0.45 arcsec and a same structure in successive pairs of back-to-back outflows. This unexpected result raises two questions: what is precisely causing the effect? And how far away from the star can the wind morphology be reliably evaluated?

\begin{figure*}
  \centering
  \includegraphics[width=4.2cm,trim=0.cm 1.cm 1.cm 1.cm,clip]{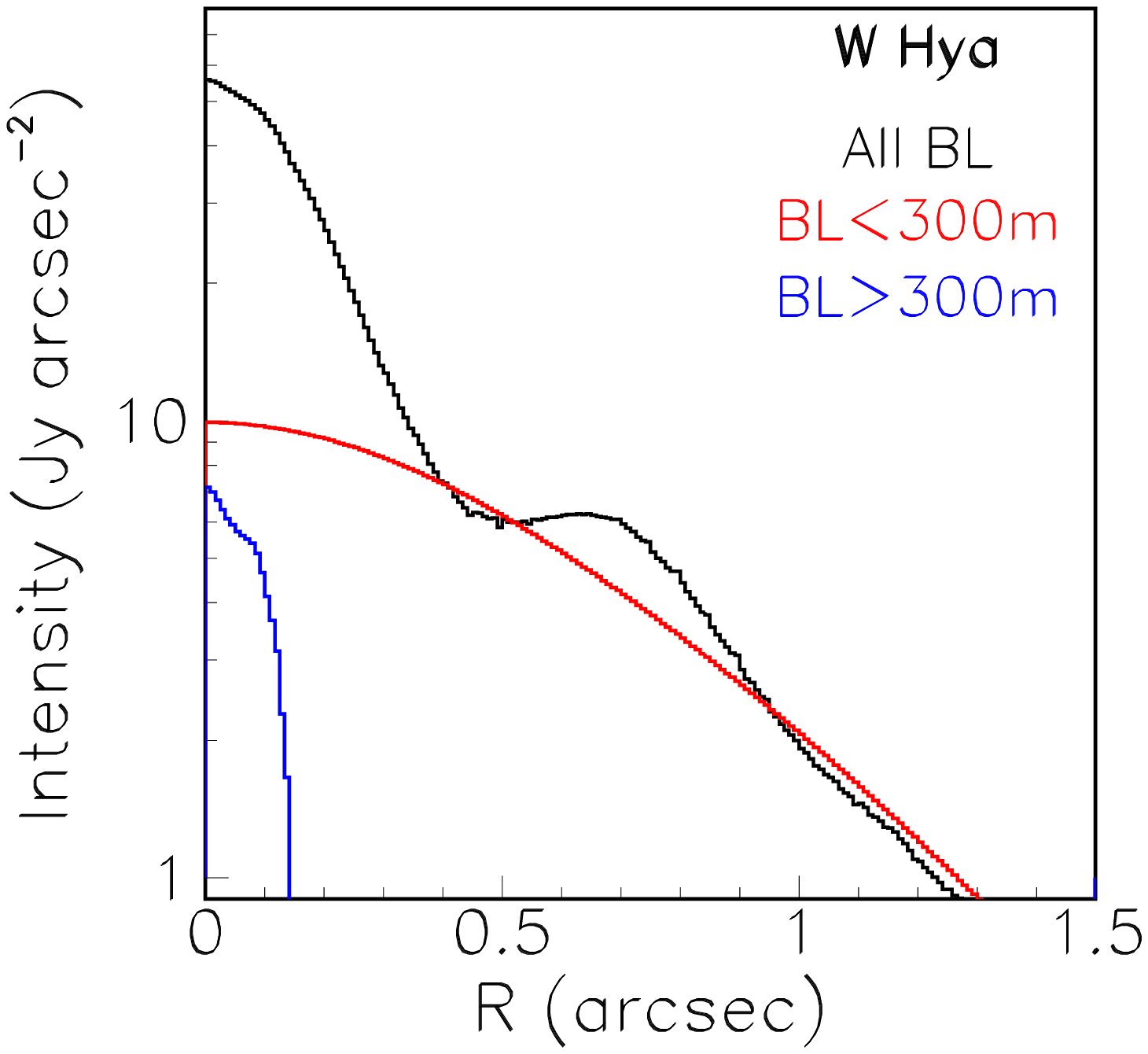}
  \includegraphics[width=4.2cm,trim=0.cm 1.cm 1.cm 1.cm,clip]{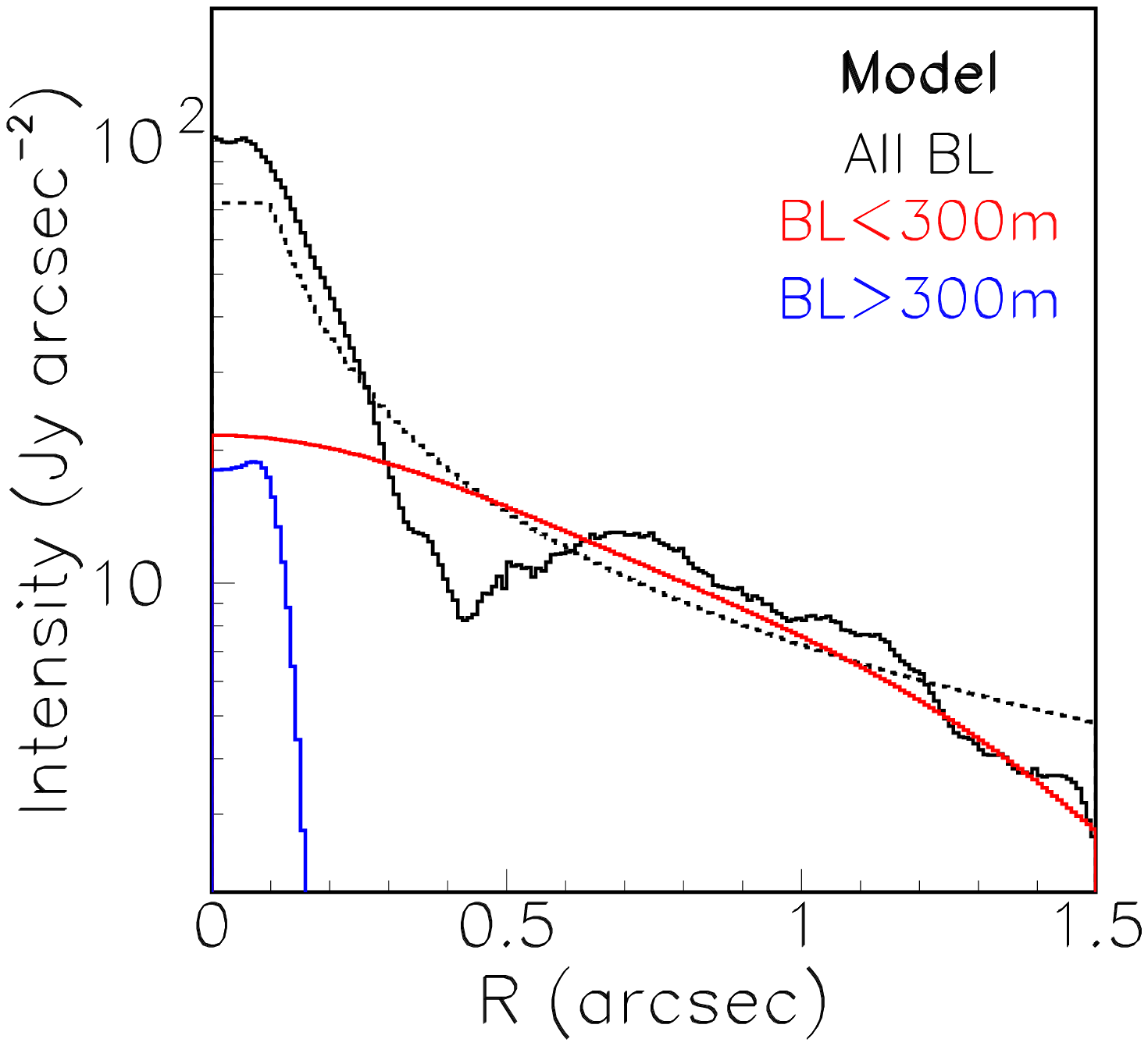}
  \includegraphics[width=4.2cm,trim=0.cm 1.cm 1.cm 1.cm,clip]{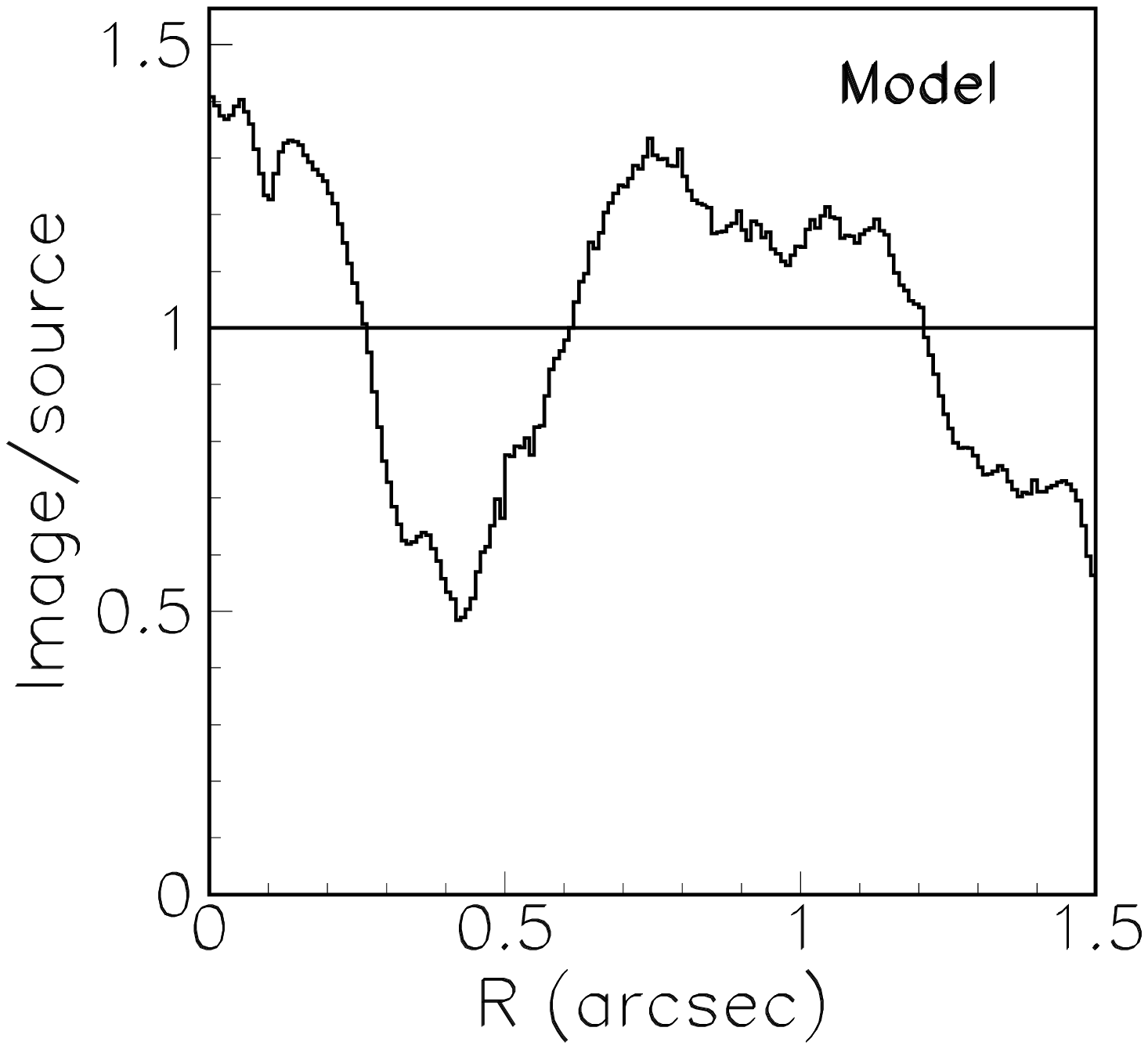}
  \includegraphics[width=4.2cm,trim=0.cm 1.cm 1.cm 1.cm,clip]{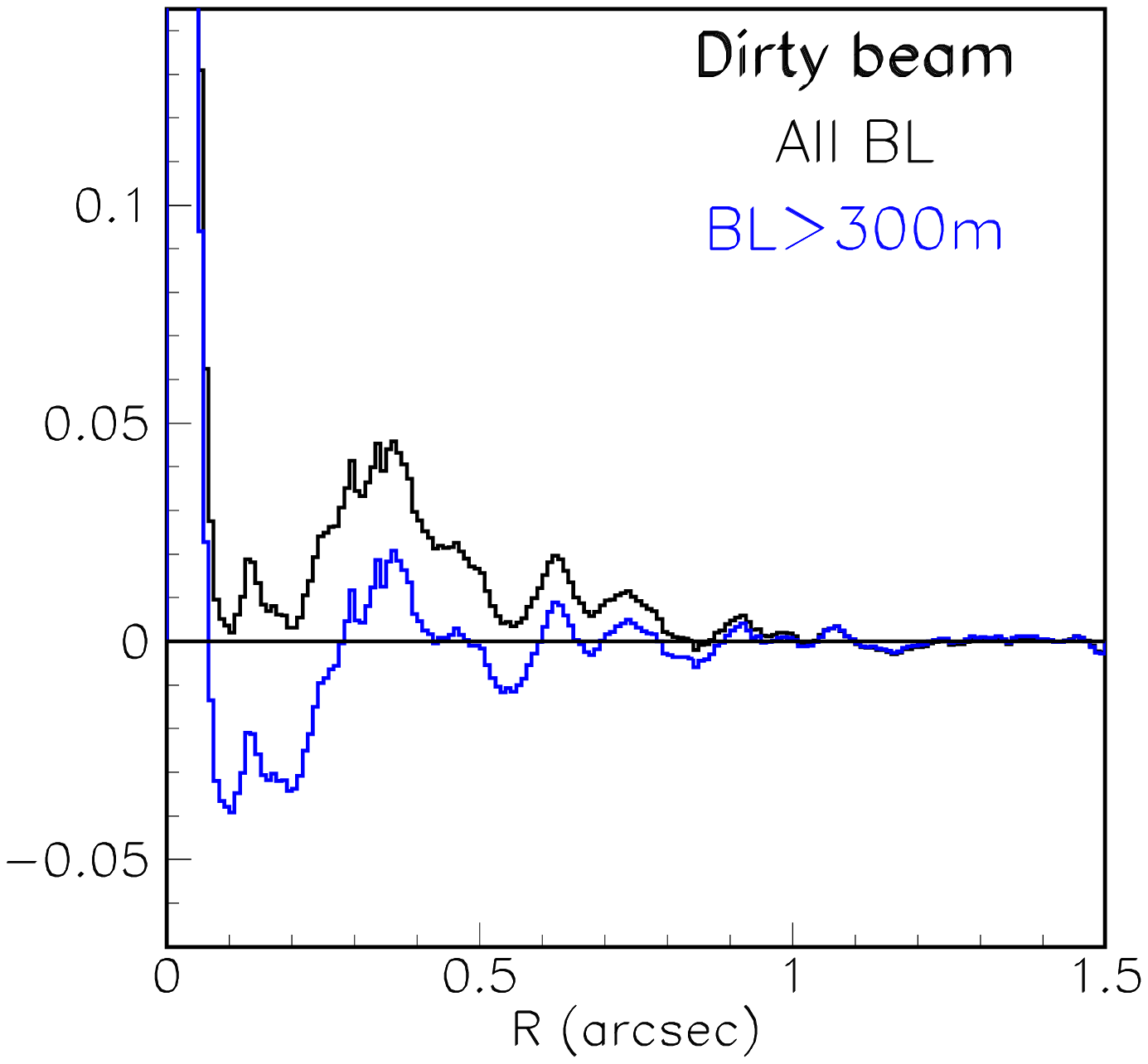}
  \caption{Left and middle-left panels: Radial distributions averaged over position angle obtained from the clean images (natural weighting) when using the whole array (black), the central array (red) or the extended array (blue). The left panel is for the $^{29}$SiO(8-7) emission averaged over $|V_z|$$<$5 km s$^{-1}$ and the middle-left panel for the isotropic wind model, shown as a dotted line. The separation between central and extended arrays is made by a simple cut at a baseline length of 300 m. We have checked that a same result is obtained by retaining instead the proper antennas. We have checked that a point source (quasar) produces a same flux density in the clean maps associated with each antenna configuration. Middle-right panel: radial dependence of the ratio between the source brightness and its image (natural weighting) for the isotropic wind model. Right panel: radial profiles of the dirty beams (averaged over position angle) of the whole array (black) and the extended array (blue). The main lobe is cut in order to properly show the wings of the profiles.}
 \label{fig6}
\end{figure*}

The left panel of Figure \ref{fig6} displays the respective contributions of the central and extended arrays to the radial brightness distribution of the $^{29}$SiO(8-7) emission. Cleaning uses natural weighting. The former is seen to probe large distances from the star and the latter short distances. The associated beam sizes (FWHM) are  63$\times$52, 58$\times$46 and 1100$\times$880 mas$^2$, for the whole, extended and central arrays, respectively. The middle-left panel shows the same distributions for the isotropic wind model described above. The central array is again seen to probe large distances and the extended array, short distances.  In this case we can compare (right panel) the real radial distribution of the source brightness with its image obtained from the whole array: it shows a broad depression at the scale of $\pm$0.2 arcsec, centred at an approximate distance of 0.45 arcsec: the missing baselines between 200 and 400 m are shown to cause an oscillation of the detected brightness about the source brightness with a peak to valley ratio of $\sim$2.6.  This oscillation is seen at a similar level on the data (left panel of Figure \ref{fig6} and Figure 2b of \citealt{Takigawa2017}), implying that it is an artefact. 

In order to shed light on its genesis, we display in the right panel of Figure \ref{fig6} the radial profiles of the dirty beams associated with the whole and extended arrays, respectively. The profile of the beam of the extended array oscillates beyond the main lobe, the width of which is defined by the longest baselines. Qualitatively, it behaves as an Airy function but the first negative oscillation is deeper; the reason is the missing baselines below some 400 m, associated with a main lobe of width $\sim$1.3$\times\lambda$/BL $\sim$0.6 arcsec (twice the maximal recoverable scale defined in the ALMA technical handbook), which needs to be subtracted from the profile of the dirty beam. The profile of the beam of the whole array is obtained from the profile of the beam of the extended array by adding the contribution of the central array, confined below some 200 m instead of 400 m, namely associated with a main lobe twice wider than required for a proper $uv$ coverage, $\sim$1.2 arcsec. As a result, the radial profile of the dirty beam stays positive up to nearly 1 arcsec (this was already apparent on its map displayed in the upper-right panel of Figure \ref{fig1}) and the dirty map is inflated in the angular region between the main lobes of a 400 m and a 200 m baseline distributions, namely between $\sim$0.6 arcsec and $\sim$1.2 arcsec. The cleaning process reduces this excess to an oscillation about the source brightness.

None of the numbers quoted in this discussion is precisely defined: the fall of the baseline distribution of the central array covers between 100 and 200 m while the rise of the baseline distribution of the extended array covers between 400 m and 500 m.  

A consequence of the results displayed in Figure \ref{fig6} is that the brightness detected at projected distances from the star in excess of some 0.2 arcsec is obtained exclusively from the central array. The contribution of the extended array is confined to short distances from the star, the maximal recoverable scale (as defined in the ALMA technical handbook) associated with a 400 m minimal baseline being only 0.6$\times\lambda$/BL=0.27 arcsec: the role of the extended array is to provide a high angular resolution at short distances from the star. Note that using robust weighting rather than natural weighting gives a stronger weight to the extended array in comparison with the central array: as a result, as shown in Figure \ref{fig4}, the depression is attenuated and less missing flux is recovered at large projected distances from the star. 

\subsection{W Hya observations: production of artefacts}
\begin{figure*}
  \centering
  \includegraphics[width=5.5cm,trim=0.cm 1.6cm 0.2cm 1.9cm,clip]{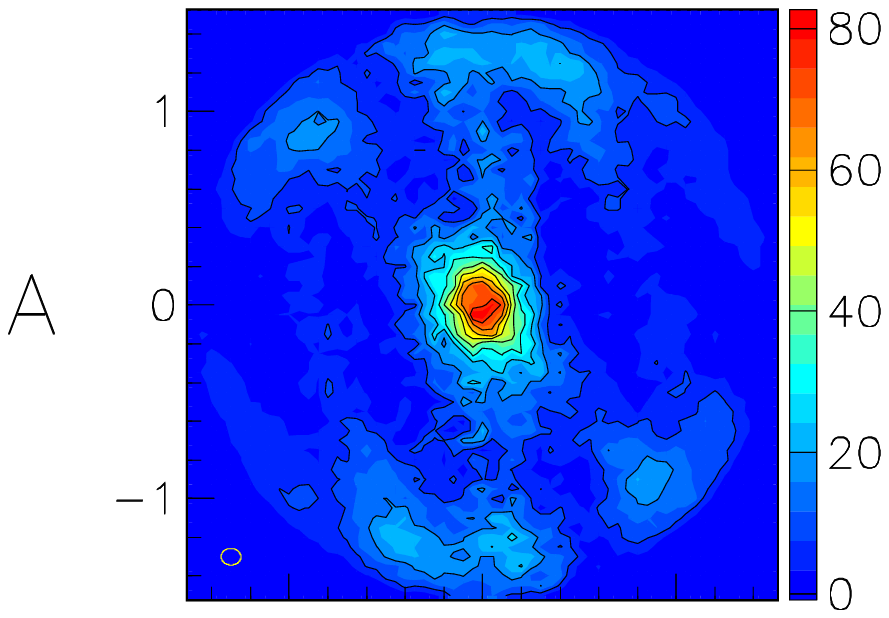}
  \includegraphics[width=5.cm,trim=-1.cm 1.6cm 0.2cm 1.9cm,clip]{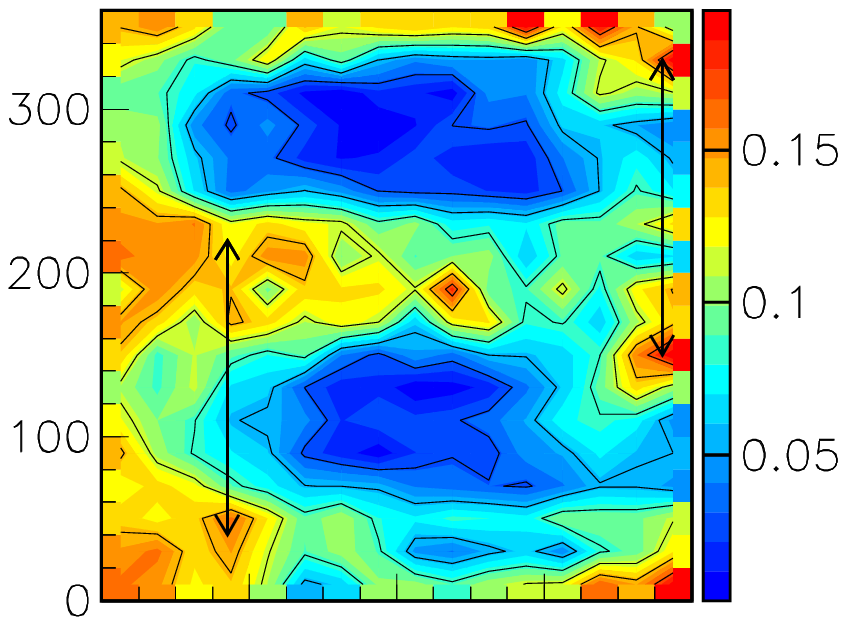}
  \includegraphics[width=5.cm,trim=-1.cm 1.6cm 0.2cm 1.9cm,clip]{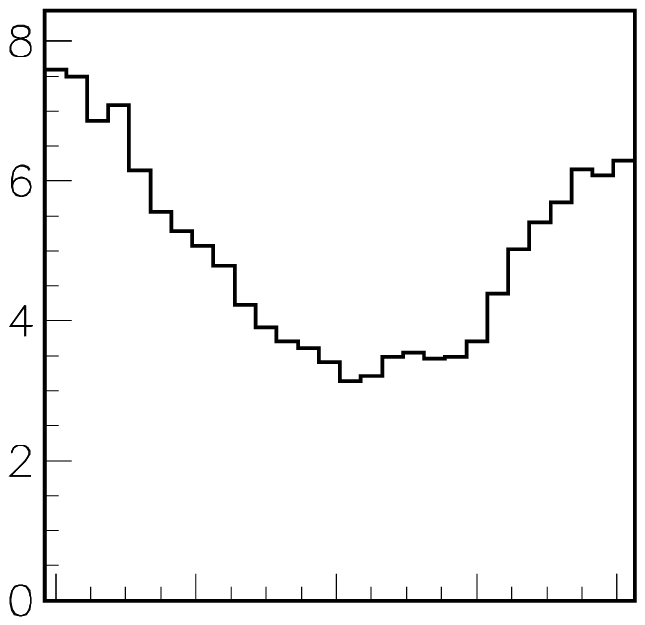}\\
  \includegraphics[width=5.5cm,trim=0.cm 1.6cm 0.2cm 1.9cm,clip]{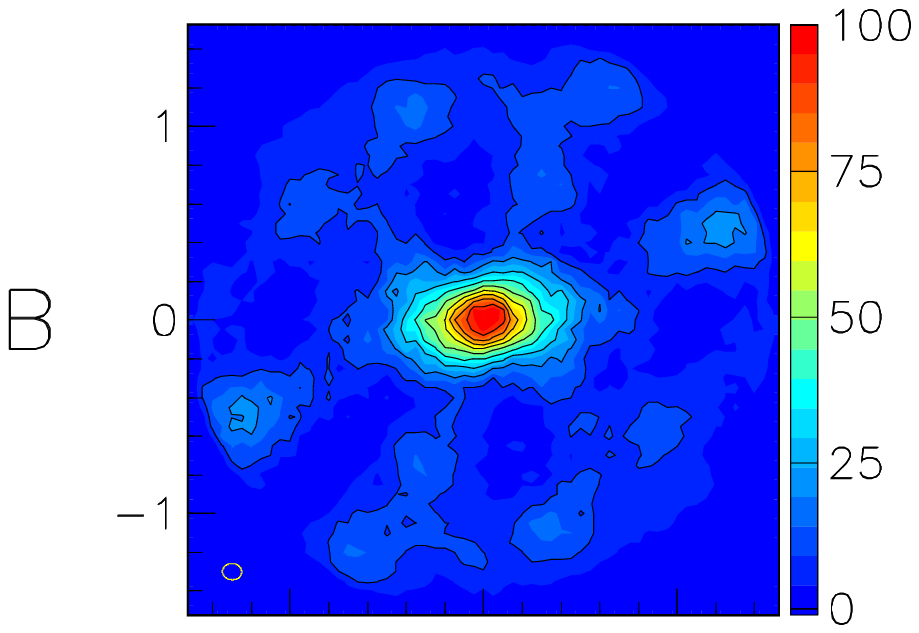}
  \includegraphics[width=5.cm,trim=-1.cm 1.6cm 0.2cm 1.9cm,clip]{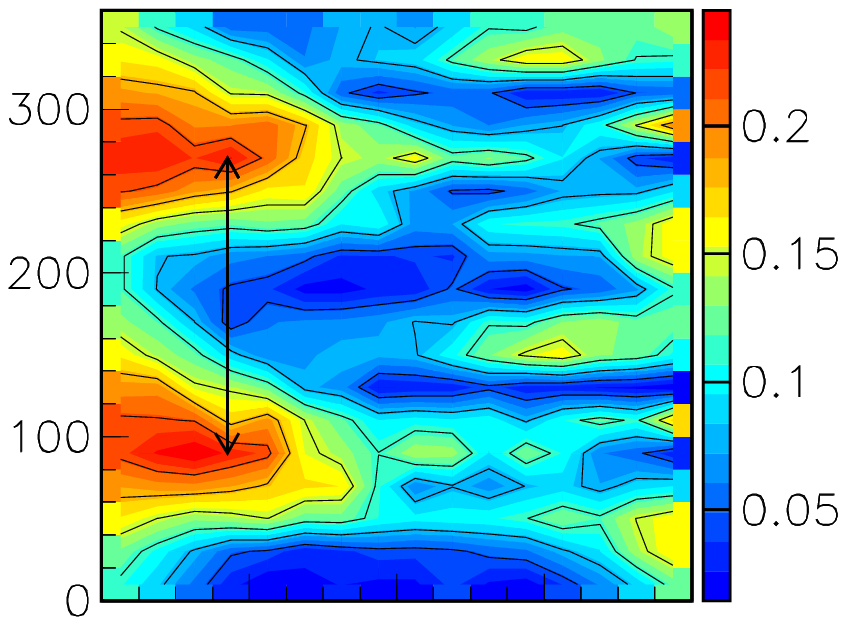}
  \includegraphics[width=5.cm,trim=-1.cm 1.6cm 0.2cm 1.9cm,clip]{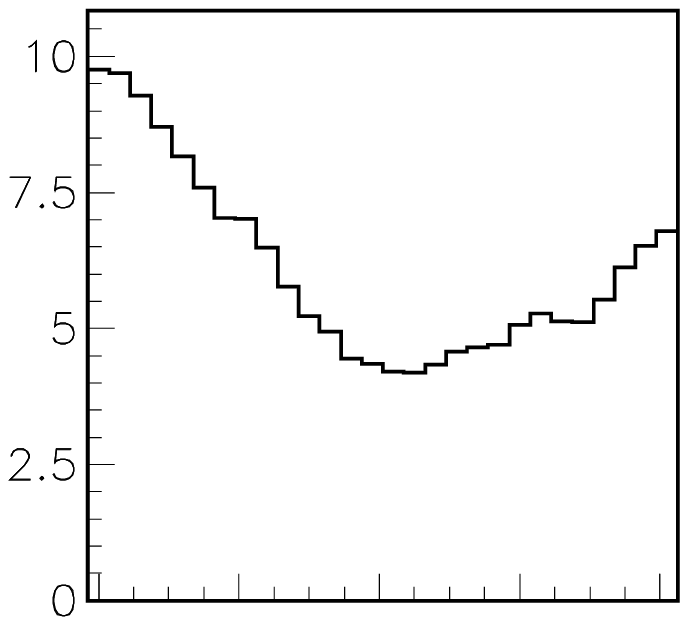}\\
  \includegraphics[width=5.5cm,trim=0.cm 1.6cm 0.2cm 1.9cm,clip]{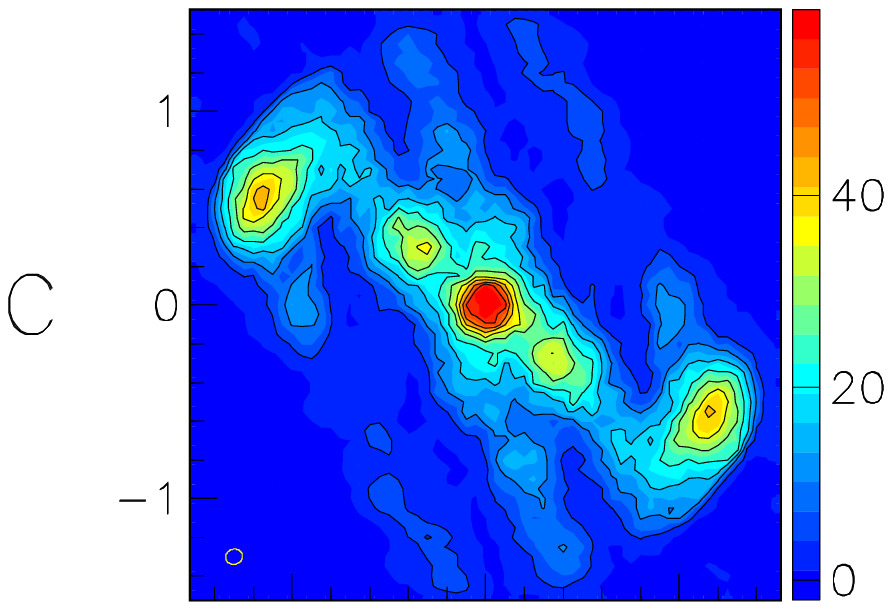}
  \includegraphics[width=5.cm,trim=-1.cm 1.6cm 0.2cm 1.9cm,clip]{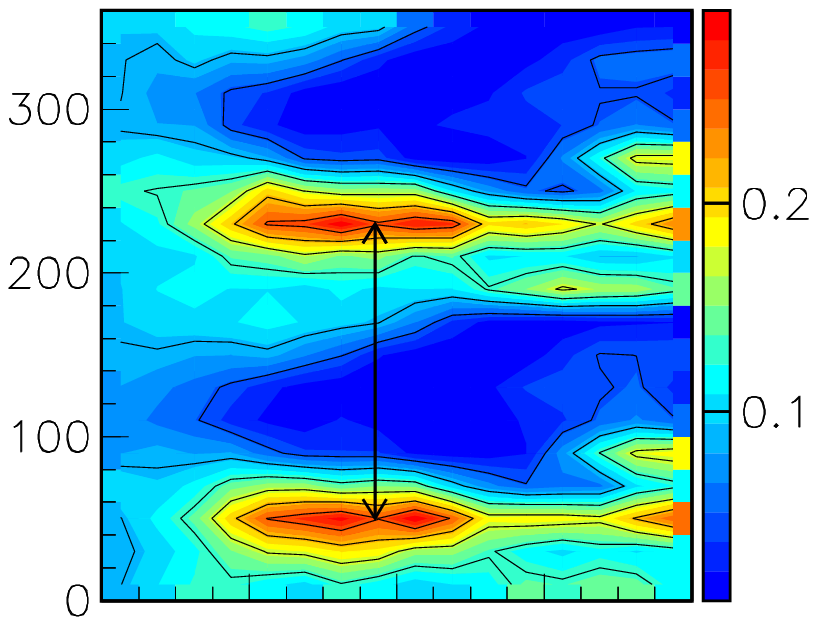}
  \includegraphics[width=5.cm,trim=-1.cm 1.6cm 0.2cm 1.9cm,clip]{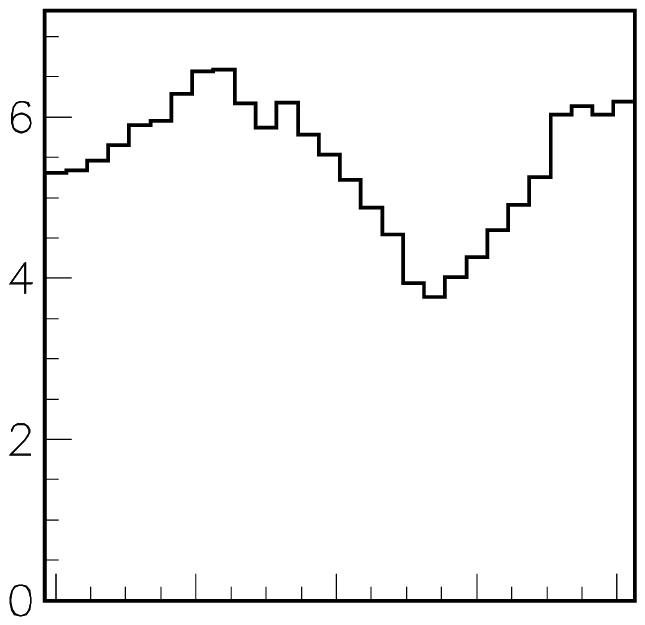}\\
  \includegraphics[width=5.5cm,trim=0.cm 1.6cm 0.2cm 1.9cm,clip]{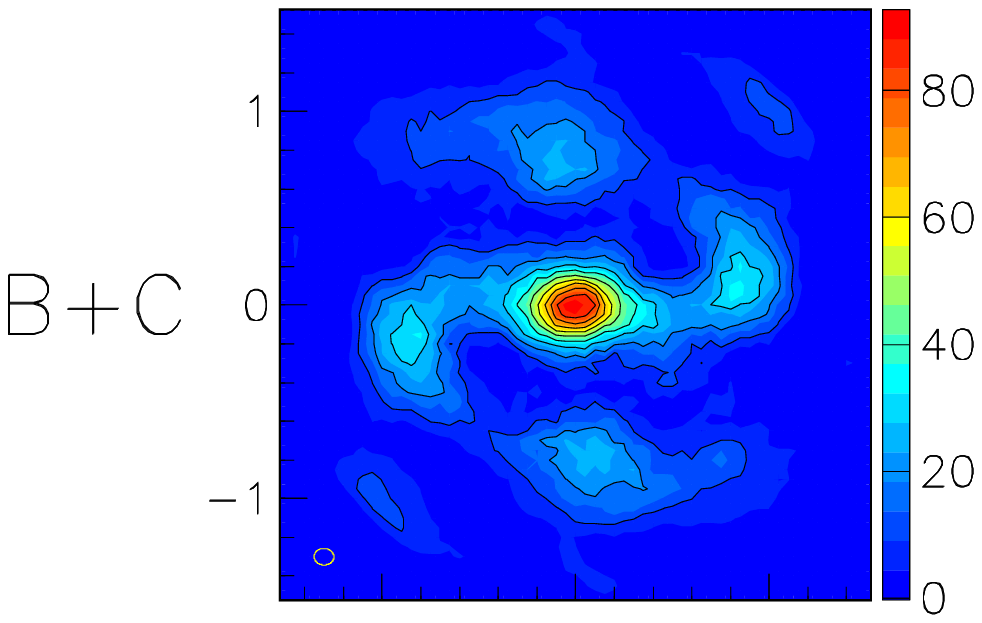}
  \includegraphics[width=5.cm,trim=-1.cm 1.6cm 0.2cm 1.9cm,clip]{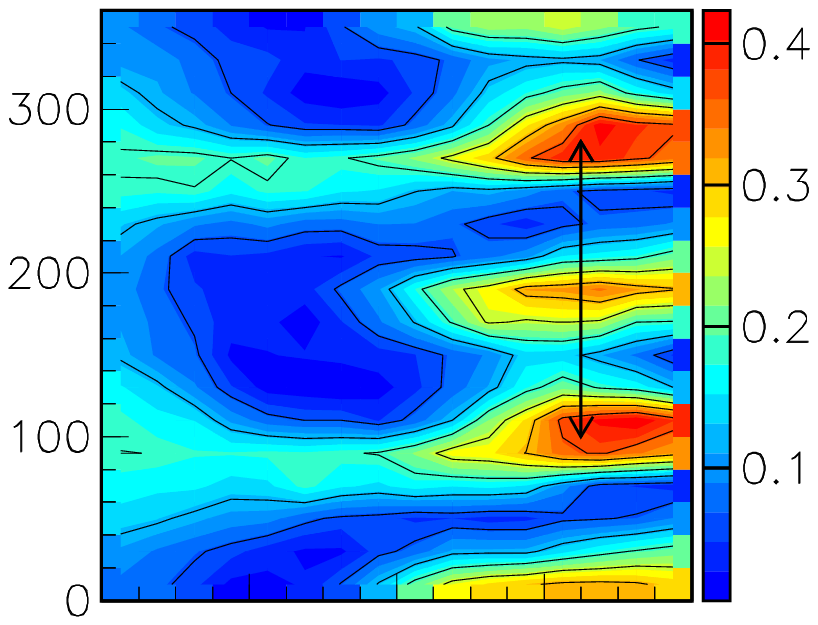}
  \includegraphics[width=5.cm,trim=-1.cm 1.6cm 0.2cm 1.9cm,clip]{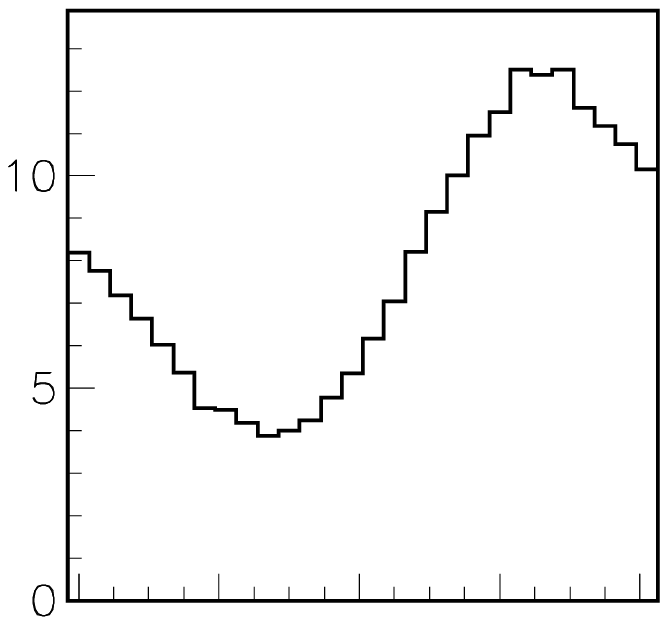}\\
  \includegraphics[width=5.5cm,trim=0.cm 1.6cm 0.2cm 1.9cm,clip]{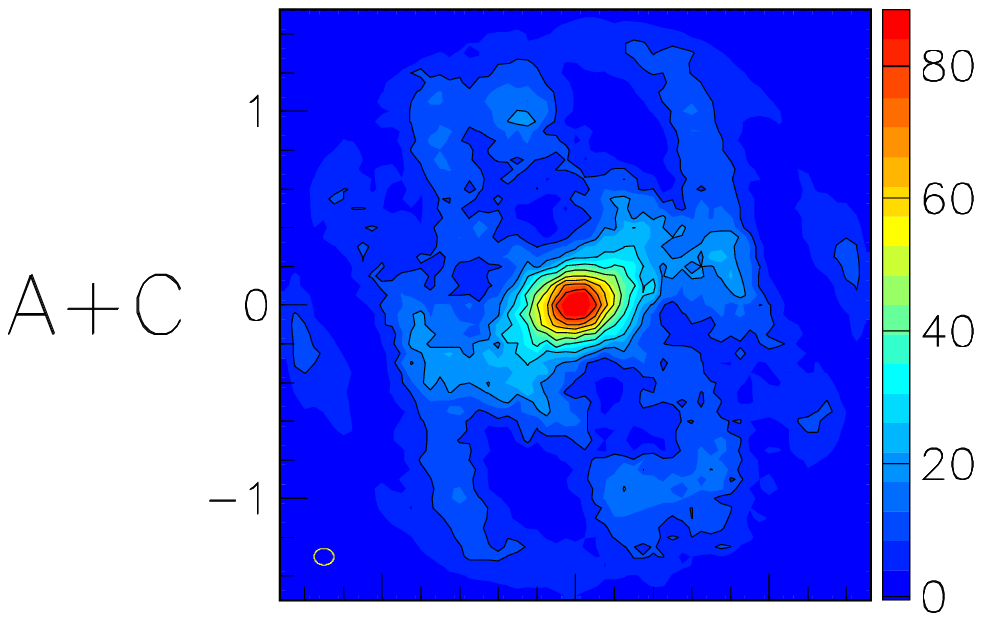}
  \includegraphics[width=5.cm,trim=-1.cm 1.6cm 0.2cm 1.9cm,clip]{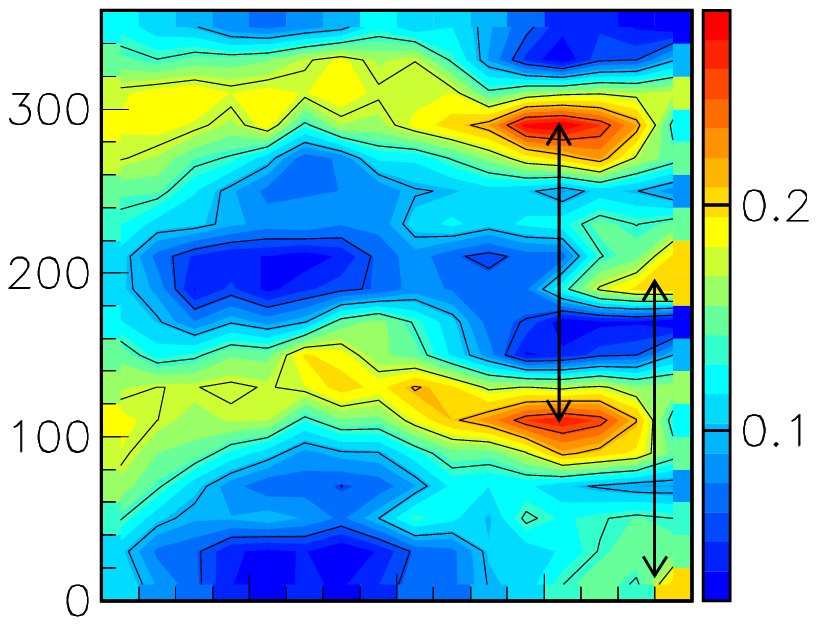}
  \includegraphics[width=5.cm,trim=-1.cm 1.6cm 0.2cm 1.9cm,clip]{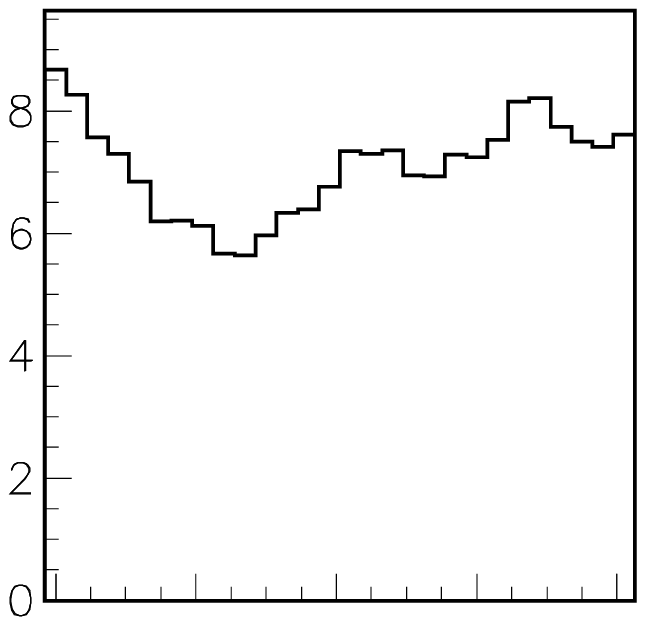}\\
  \includegraphics[width=5.5cm,trim=0.cm 1.6cm 0.2cm 1.9cm,clip]{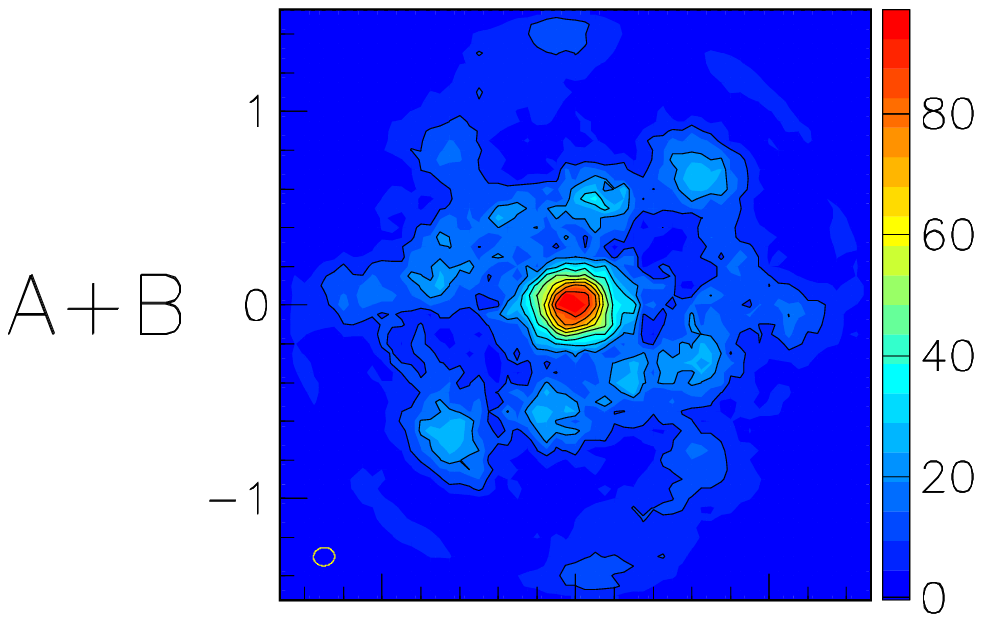}
  \includegraphics[width=5.cm,trim=-1.cm 1.6cm 0.2cm 1.9cm,clip]{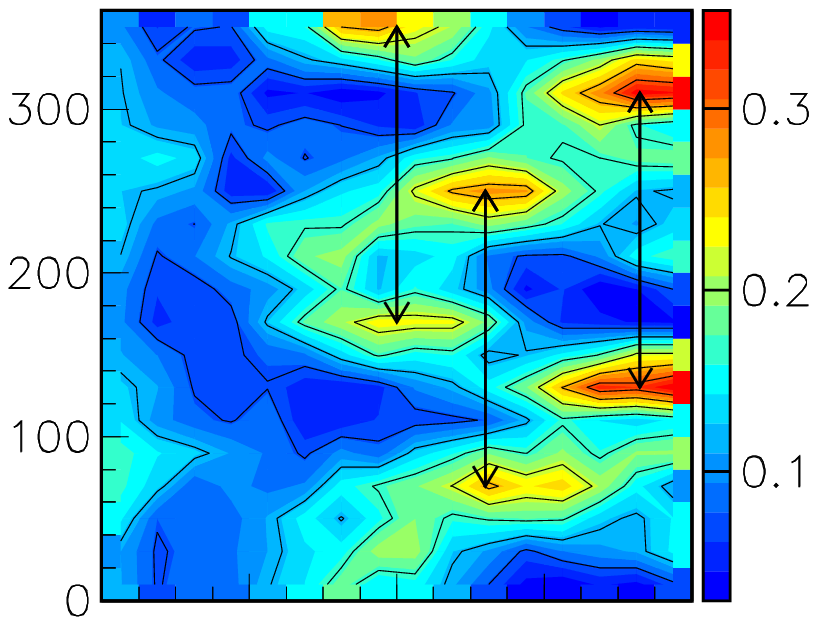}
  \includegraphics[width=5.cm,trim=-1.cm 1.6cm 0.2cm 1.9cm,clip]{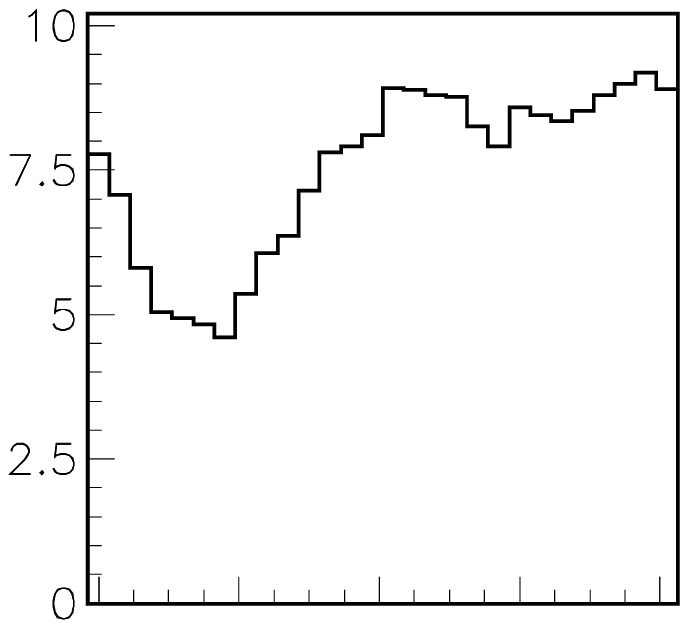}\\
  \includegraphics[width=5.5cm,trim=0.cm 0.5cm 0.2cm 1.9cm,clip]{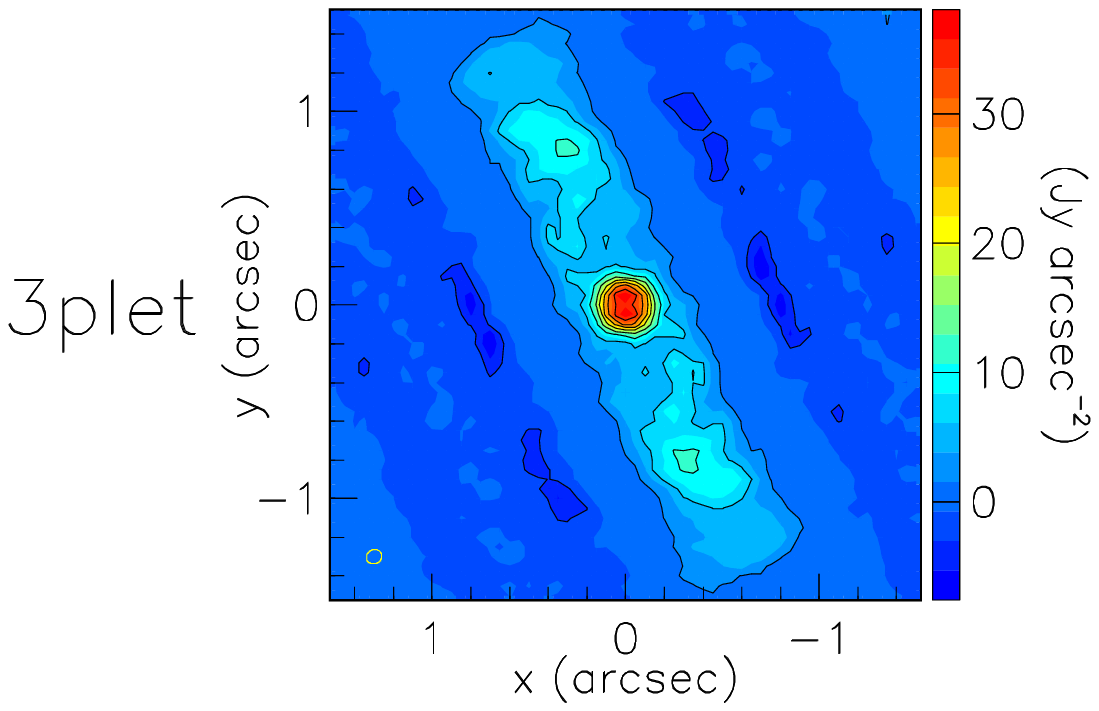}
  \includegraphics[width=5.cm,trim=-1.cm 0.5cm 0.2cm 1.9cm,clip]{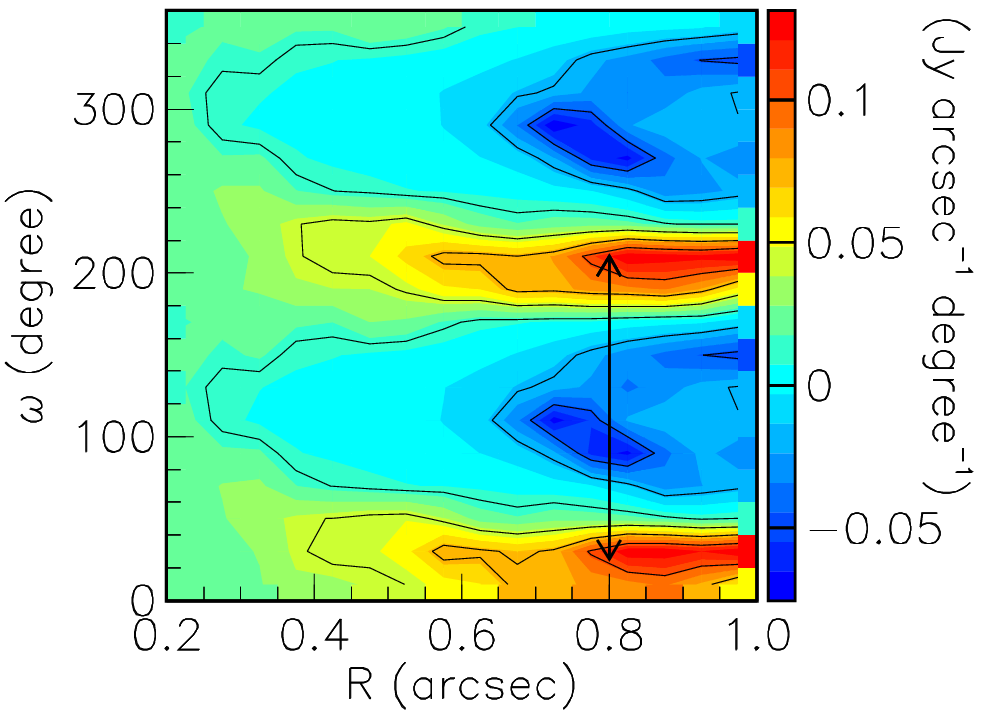}
  \includegraphics[width=5.cm,trim=-1.cm 0.5cm 0.2cm 1.9cm,clip]{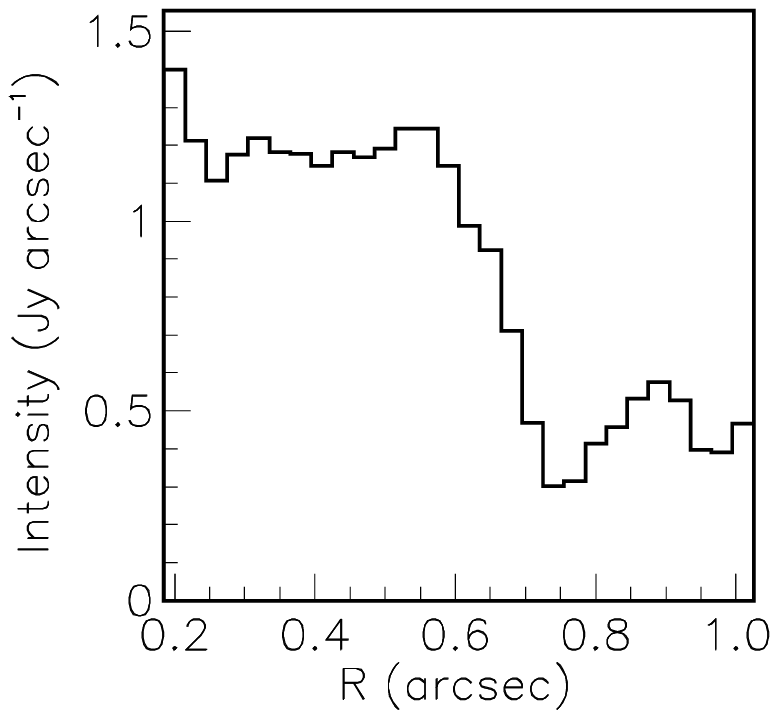}

  \caption{Clean maps of the isotropic wind model emission obtained by using the whole extended array but retaining only a subset of antennas in the  central array. Intensity maps are shown on the left column in Cartesian coordinates ($y$ vs $x$) and in the middle column in polar coordinates ($\omega$ vs $R$); the radial distributions, integrated on position angle, are shown in the right column. From up down, the subsets are A, B, C, B+C, A+C and A+B where A, B and C are defined in the central lower panel of Figure \ref{fig1}. The last row ignores all antennas of the central array, the pattern being exclusively defined by the triplet of antennas indicated by a blue arrow in Figure \ref{fig1}.}
 \label{fig7}
\end{figure*}

The arguments developed in the preceding sub-section have clarified the role of the missing baselines in producing a depression of the radial distribution of the detected brightness covering projected distances from the star between approximately 0.2 and 0.6 arcsec. They have not addressed the production of artefacts mimicking the emission of back-to-back outflows, evidence for which is given in the left panel of Figure \ref{fig5}, but they have shown that the image obtained at projected distances from the star in excess of $\sim$0.2 arcsec is governed by the antenna configuration of the central array. Strong additional evidence is shown in Figure \ref{fig7} by displaying images obtained with selected subsamples of the antennas of the central array. A particularly spectacular illustration is obtained by disregarding the central array altogether (lower row of the figure). Then short baselines are provided exclusively by the triplet of antennas indicated by a blue arrow in Figure \ref{fig1}. They completely dominate the pattern observed beyond $R\sim$0.2 arcsec, which is now a set of fringes oriented at position angle 25\dego\ modulo 180\dego\ and separated by 0.6 arcsec.
 
\begin{figure*}
  \centering
  \includegraphics[height=4.8cm,trim=0.0cm 1.cm 0.cm 0cm,clip]{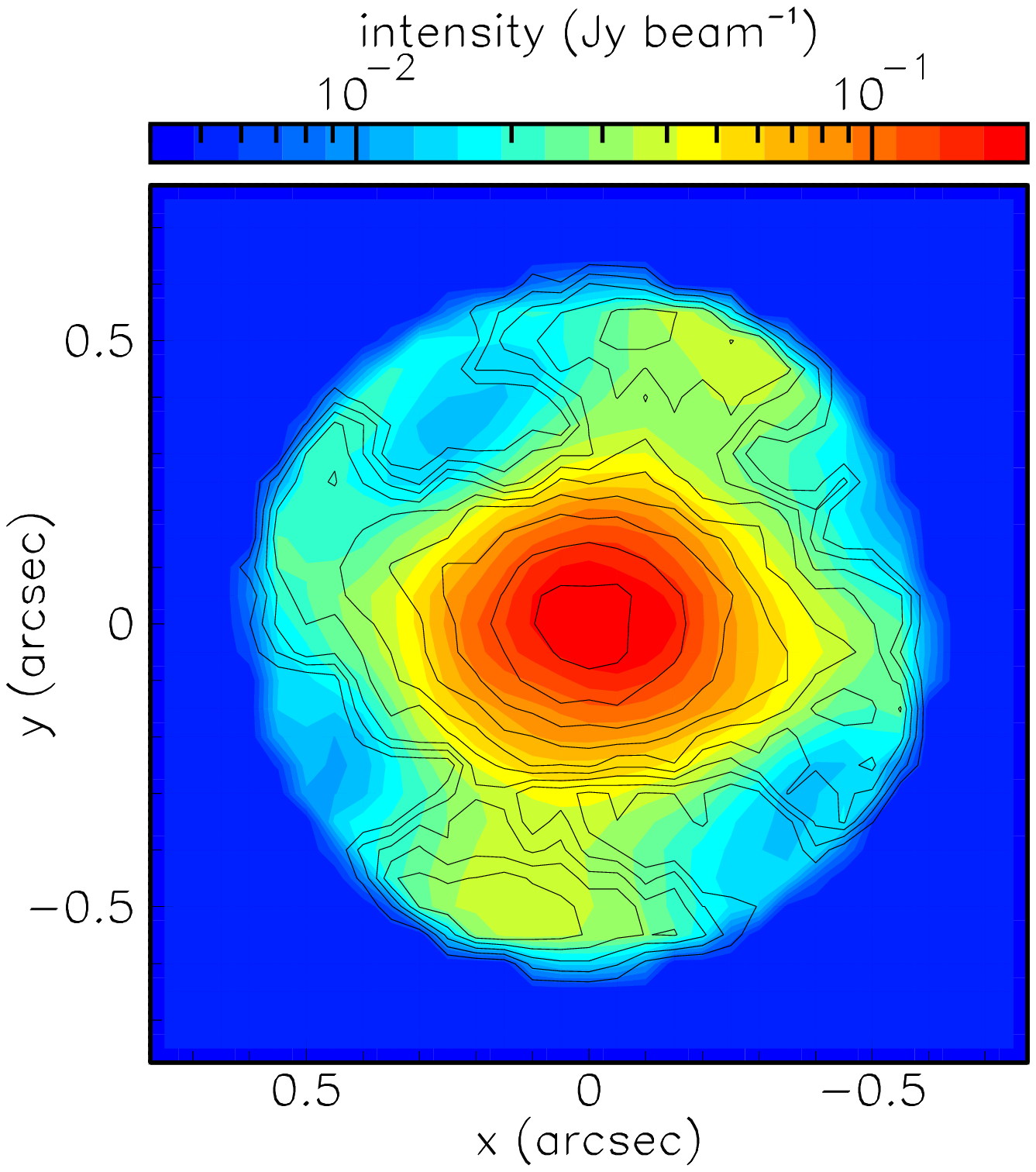}
  \includegraphics[height=4.8cm,trim=0.0cm 1.cm 0.cm 0cm,clip]{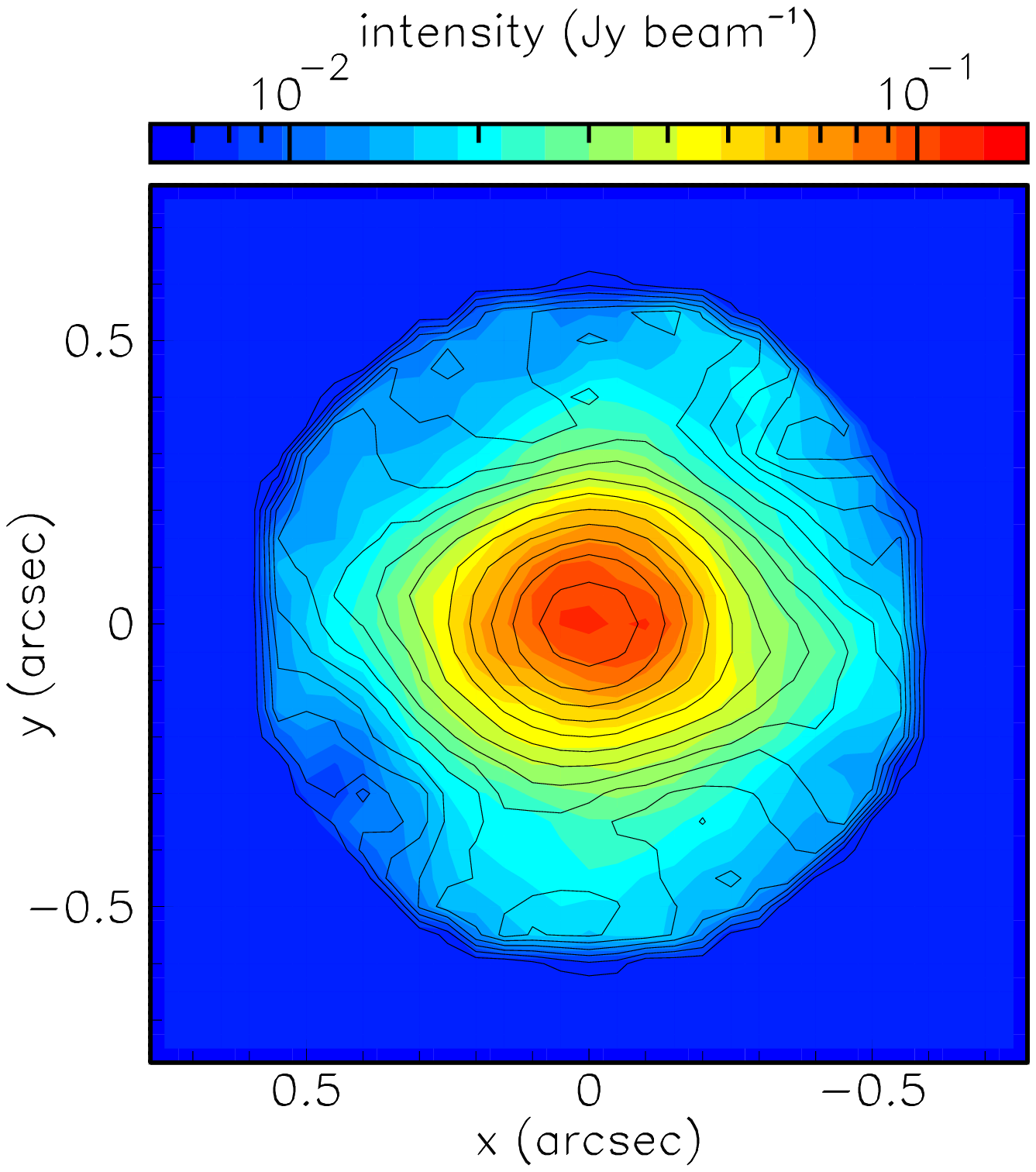}
  \includegraphics[height=4.8cm,trim=0.0cm 1.cm 0.cm 0cm,clip]{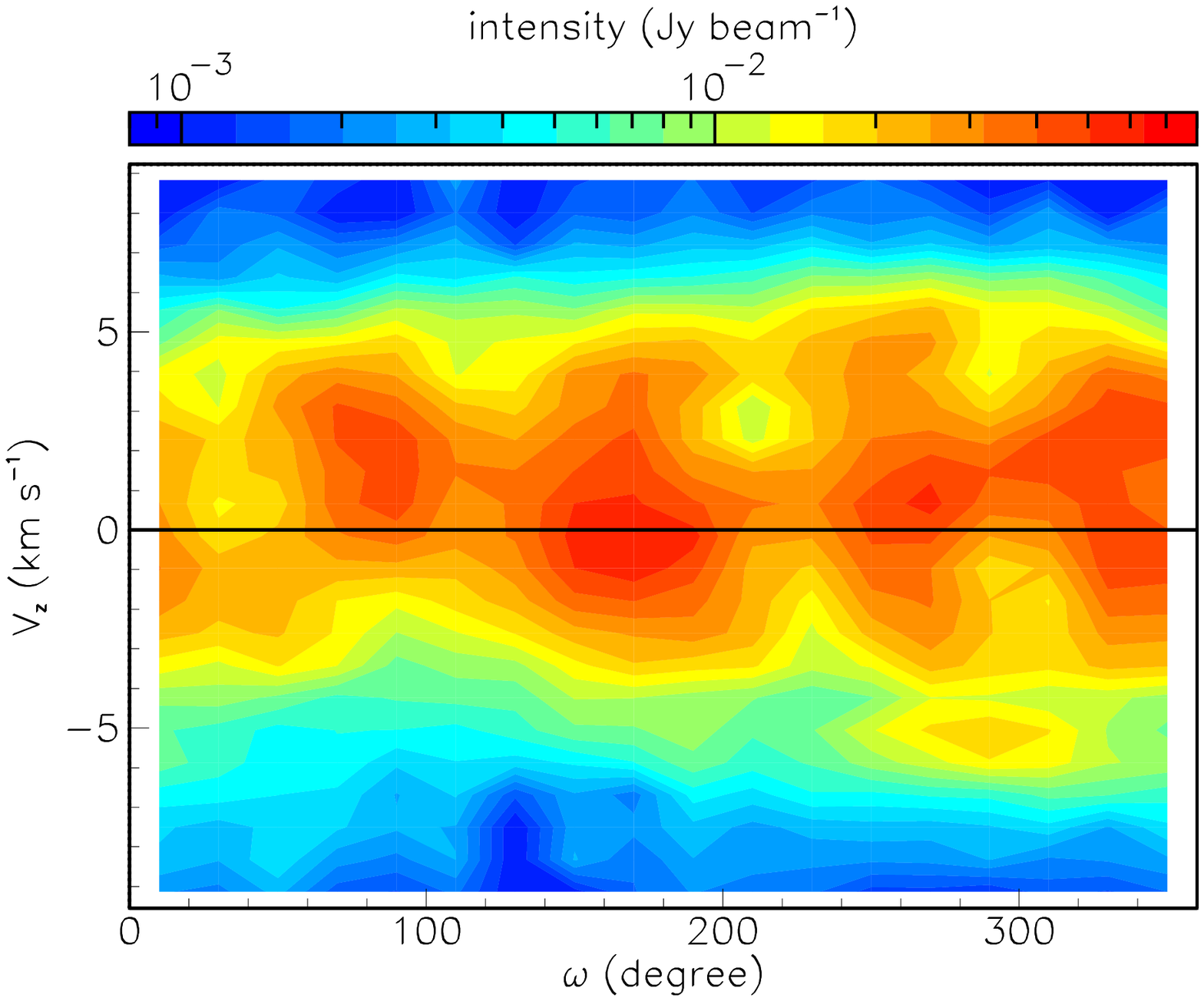}
  \caption{Left and centre: comparison between the intensity map of the $^{29}$SiO emission averaged between $-$4 and 6 \kms\ (colour scale) and the model (contours). The left panel uses natural weighting and the central panel uses robust weighting. Both maps are limited to the region $R$$<$0.6 arcsec. Right:  PV map of the $^{29}$SiO emission ($V_z$ $vs$ $\omega$) averaged over the interval 0.3$<$$R$$<$0.6 arcsec.}
 \label{fig8}
\end{figure*}

While the missing baselines are to be blamed for the production of the radial depression, the production of artefacts at larger projected distances from the star is therefore to be blamed exclusively on the insufficient $uv$ coverage provided by the central array. A central array offering a better $uv$ coverage would have made it possible to obtain reliable images up to projected distances from the star at arcsec scale. In the present case, however, this is unrealistic. The left and central panels of Figure \ref{fig8} compare the images of the data and of the model in projection on the sky plane, showing strong similarity between the patterns displayed by the two images: qualitatively, it suggests that the emission of the W Hya CSE is isotropic up to distances of at least 0.5 arcsec from the centre of the star. However, ascertaining such a statement requires new observations providing adequate coverage of the $uv$ plane. 

The above discussion shows the importance of ensuring that the detection of back-to-back outflows is not in fact the result of artefacts caused by an improper $uv$ coverage. Such pairs of outflows have been observed earlier in the CSEs of several AGB stars, such as R Dor \citep{Nhung2021}, EP Aqr \citep{Nhung2019}, RS Cnc \citep{Winters2021}, etc. In all these cases a distinctive feature was the observed asymmetry of the Doppler velocities of the members of the pair. This provides indeed a useful criterion against artefacts, but cannot be used if the outflows are in the plane of the sky. The right panel of Figure \ref{fig8} displays the $V_z$ vs $\omega$ map of the emission of the $^{29}$SiO line averaged in the region 0.3$<$R$<$0.6 arcsec where the artefacts are apparent. It shows the absence of significant back-to-back asymmetry in $V_z$. We checked that this result is independent from the specific $R$ interval used to make the comparison. We note the presence of a small blob of enhanced emission centred at $V_z$$\sim$5.5 km s$^{-1}$ and $\omega\sim$290\dego, which we discuss briefly at the end of the next section.

\section{W Hya near the stellar disc: a comparison with R Dor}{\label{sec4}}
The arguments developed in the preceding sections have shown that the observed data-cube of $^{29}$SiO(8-7) emission is reliable up to distances of over  15 au (0.15 arcsec) from the star. In this range, W Hya has recently been the object of detailed studies: observations in the visible and near-visible using NACO and SPHERE-ZIMPOL on the VLT and AMBER on the VLTI \citep{Norris2012, Ohnaka2016, Ohnaka2017, Khouri2020, Hadjara2019} have given evidence for a clumpy and dusty layer, displaying important variability at short time scale. Dust grains, mostly aluminium composites, are found to have sizes ranging between 0.1 and 0.5 microns. In the near-IR, using MIDI on the VLTI, \citet{ZhaoGeisler2015} have confirmed this result as have \citet{Vlemmings2017, Vlemmings2019} using ALMA to observe continuum and CO($\nu$=1, $J$=3-2) emissions close to the star. \citet{Khouri2015} have proposed a simple model that also confirms this result, showing that SiO emits at larger distances. \citet{Khouri2014a,Khouri2014b} have used Herschel in the infrared to measure the $^{12}$C/$^{13}$C ratio as 18$\pm$10, to detect emission from H$_2$O and $^{28}$SiO and to establish that $\sim$1/3 of SiO atoms are locked up in dust.  \citet{Takigawa2017}, in their analysis of the present observations, have shown that while the spatial distribution of AlO molecules is confined within $\sim$6 au, $^{29}$SiO molecules extend beyond $\sim$10 au without significant depletion. Takigawa et al. (2019) argue that transition alumina containing $\sim$10\% of Si is the most plausible source of the dust emission from alumina-rich AGB stars. Also using the present observations, \citet{Vlemmings2017} have explored the inner layer of the CSE. The latter two publications, by \citet{Takigawa2017} and \citet{Vlemmings2017}, using the present observations, have obtained important results concerning the shock-heated atmosphere of the star and the mechanism governing the formation of dust, respectively. This does not leave much room to extract from the present observations further information of relevance. Yet, a few features that have not been mentioned in the published literature deserve being briefly presented and commented upon, which we do in the present section. To do so, we use as a guide ALMA observations of the emission by the CSE of R Dor of the same $^{29}$SiO(8-7) molecular line as observed in W Hya \citep{Nhung2021}. The motivation for doing so is the similarity between the two stars (same mass loss rate, very similar long periods and spectral types, absence of technetium in their spetra, etc.) which is often used in the published literature as an encouragement for comparing their properties, as is the case in particular in the work of \citet{Vlemmings2017}.

\begin{figure*}
  \centering
  \includegraphics[width=4.5cm,trim=0.0cm .5cm 1.cm 0.5cm,clip]{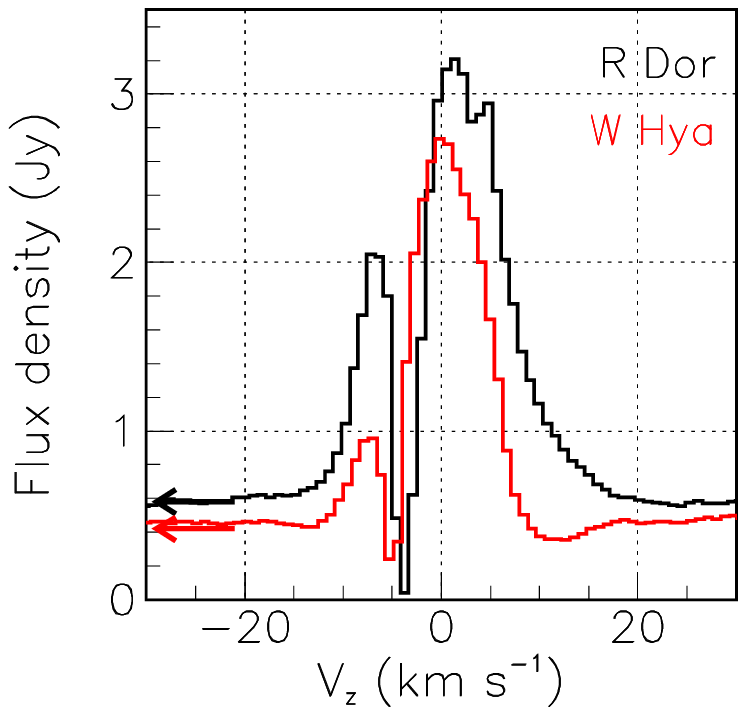}
  \includegraphics[width=4.5cm,trim=0.0cm .5cm 1.cm 0.5cm,clip]{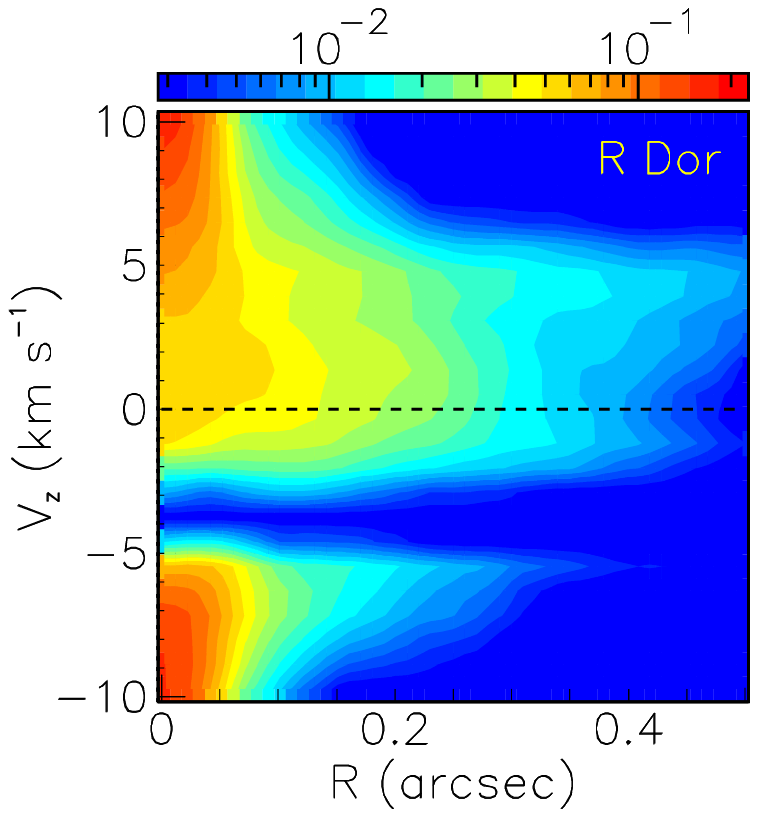}
  \includegraphics[width=4.5cm,trim=0.0cm .5cm 1.cm 0.5cm,clip]{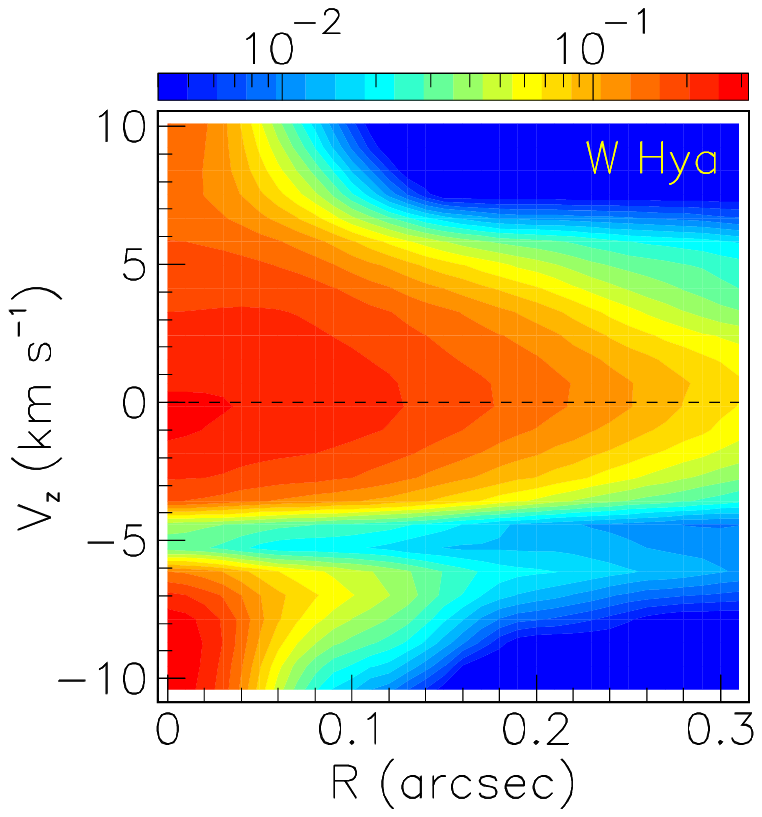}
  \caption{Doppler velocity spectra integrated over $R$$<$12 au  for R Dor (black) and W Hya (red). The arrows show the continuum levels measured by \citet{Vlemmings2019}. Centre and right: PV maps $V_z$ vs $R$ averaged over position angles, for R Dor (centre) and W Hya (right). The colour scales are in units of Jy beam$^{-1}$. }
 \label{fig9}
\end{figure*}

Absorption over and well beyond the stellar disc is illustrated in Figure \ref{fig9} and found to be qualitatively similar to that of R Dor. A narrow absorption peak is seen at $\sim$$-$5.5 \kms, slightly lower than for R Dor, corresponding to absorption in the outer SiO layer\footnote{We learned after submission of the present manuscript of a very recently published paper by \citet{Vlemmings2021} that gives evidence for CO(3-2) maser emission at the same Doppler velocity, $\sim$$-$5.5 \kms, as observed here as terminal velocity.}, which is expected to extend up to some 100 au from the star \citep{Nhung2021}. The continuum levels measured within 12 au, corresponding to a good coverage of the continuum emission, are in excellent agreement with those measured by \citet{Vlemmings2019}: 420 and 580 mJy for W Hya and R Dor respectively. The larger beam size (when measured in au) of the W Hya data causes significant smearing of the absorption. Note that the effect of absorption is visible over the whole $R$ range, irrespective of the SSP, which causes distortions of the image brightness but does not create a signal where there is no emission.

\begin{figure*}
  \centering
  \includegraphics[width=4.5cm,trim=0.0cm 1cm 1.5cm 1.5cm,clip]{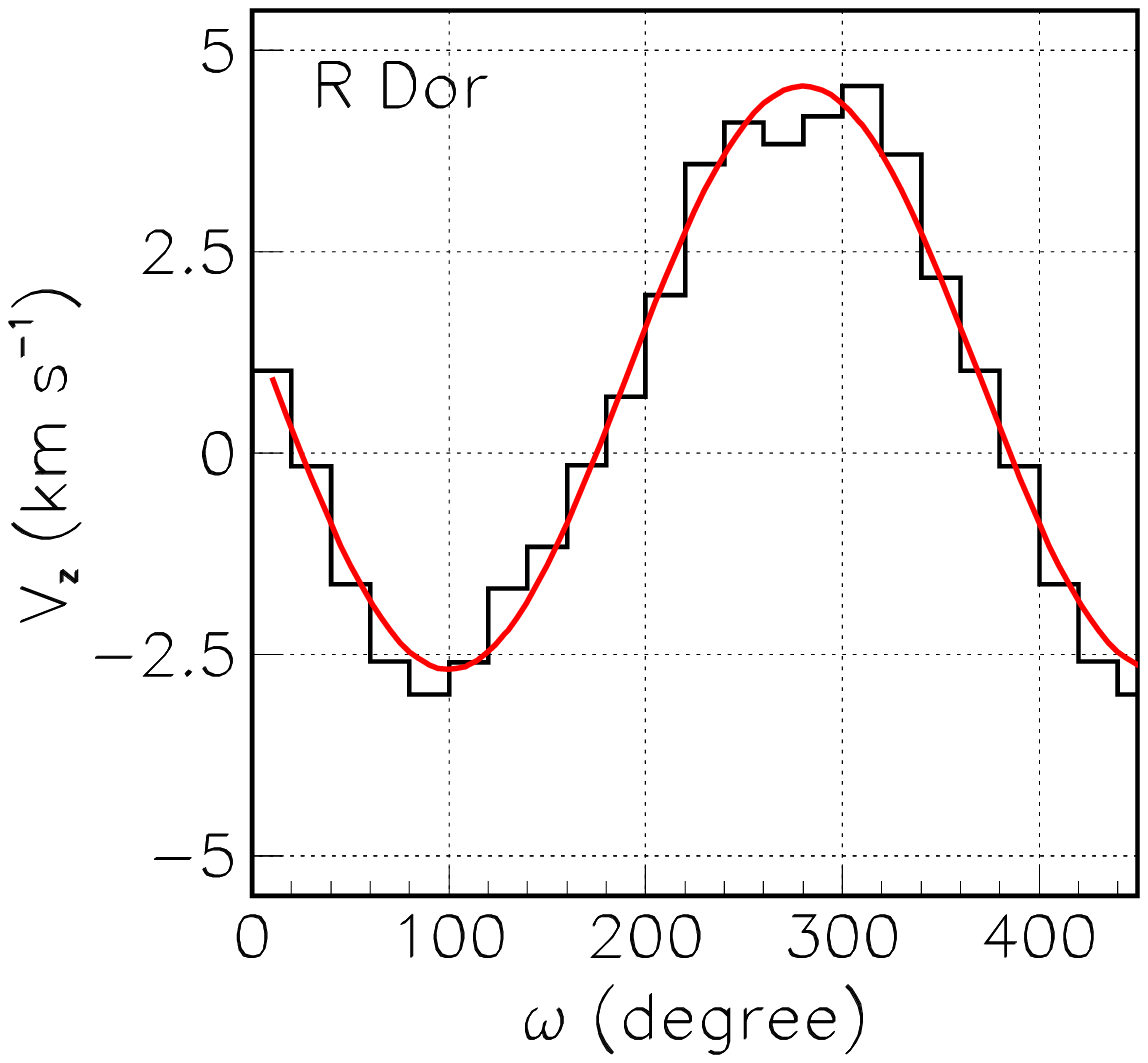}
  \includegraphics[width=4.5cm,trim=0.0cm 1cm 1.5cm 1.5cm,clip]{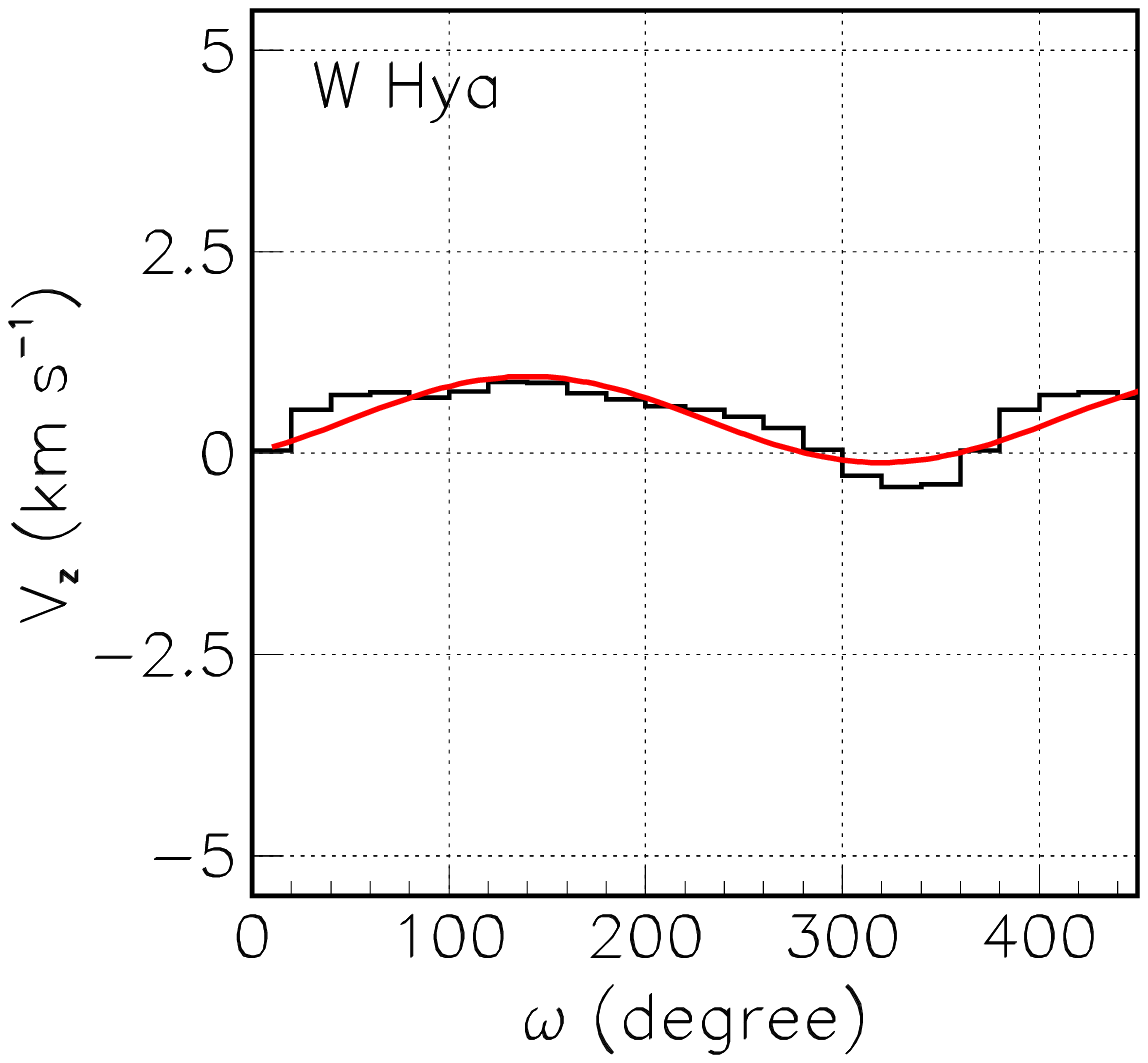}
  \caption{Dependence of $<$$V_z$$>$ (calculated over  $|V_z|$$<$12 \kms) on $\omega$ for R Dor (left) and W Hya (right) averaged in rings $R$$=$7.5$\pm$3.0 au and
$R$$=$7.5$\pm$2.5 au, respectively. The sine wave fits (\kms) are $0.5-3.6\sin(\omega-11$\dego)  and $0.4+0.5\sin(\omega-48$\dego), respectively.}
 \label{fig10}
\end{figure*}

Rotation of the R Dor CSE in the vicinity of the star has been studied in detail \citep{Vlemmings2018, Homan2018, Nhung2021} and found to be solid-body-like up to some $\sim$6 au from the star, to reach a maximal velocity of $\sim$6 \kms\ at $\sim$8 au and then to slow down and cancel beyond 15 au. It is illustrated in Figure \ref{fig10}. Instead, the W Hya observations, albeit being slightly less sensitive to the presence of a possible rotation because of the larger beam size, show no clear evidence for rotation. From the observed dependence of the mean Doppler velocity on position angle, we infer an upper limit of $\sim$1 \kms\ on a possible rotation velocity averaged between $R$$=$5 au and $R$$=$10 au and divided by the sine of the angle between rotation axis and line of sight. In both cases an offset of $\sim$0.5 \kms\ is observed, probably due, at least in part, to absorption. We note that a tilt of 5\dego\ of the rotation axis with respect to the line of sight is sufficient to prevent detection of a rotation velocity equal to that at stake in R Dor.

\begin{figure*}
  \centering
  \includegraphics[width=4.5cm,trim=0.0cm 1cm 1.5cm 1.5cm,clip]{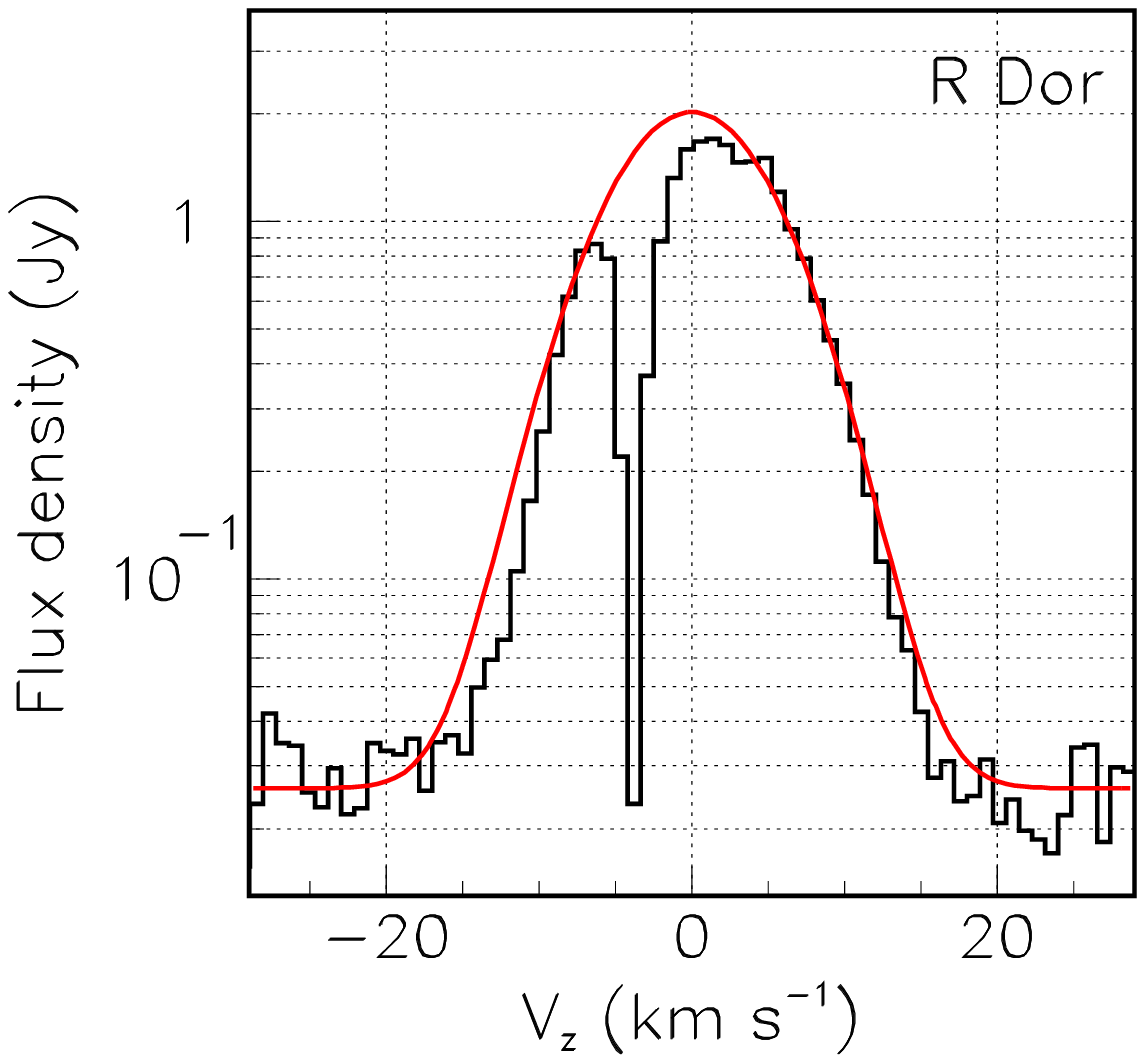}
  \includegraphics[width=4.5cm,trim=0.0cm 1cm 1.5cm 1.5cm,clip]{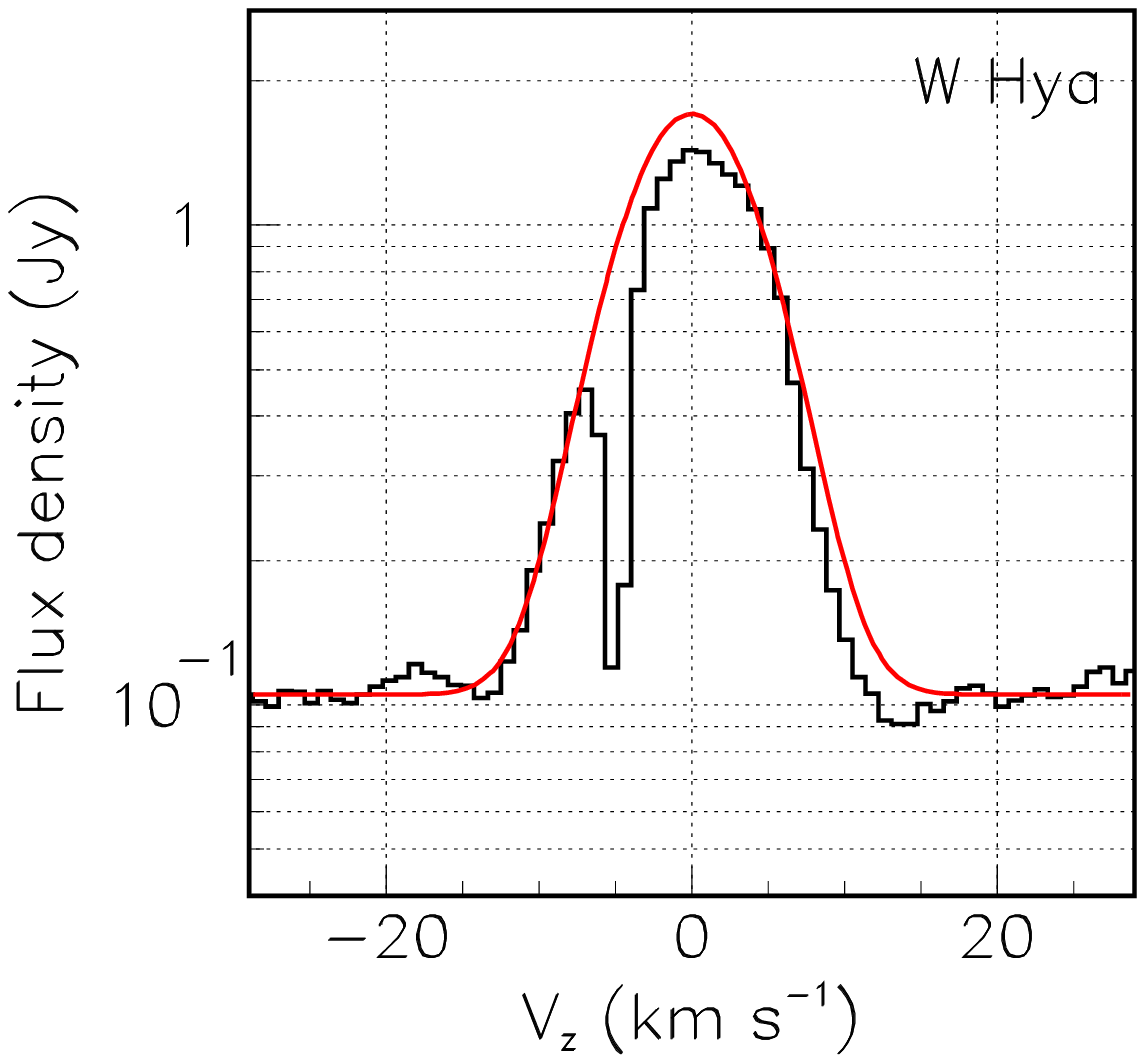}
  \caption{Doppler velocity spectra (black) integrated  in a ring of 5 to 10 au centred on R Dor (left) and W Hya (right). Gaussian profiles, superimposed on a constant flux, are shown in red as references. They have $\sigma$'s of 5.2 (left) and 4.2 (right) \kms.}
 \label{fig11}
\end{figure*}

The presence of high Doppler velocity wings in the vicinity of the star has been observed in the CSEs of several oxygen-rich AGB stars and is understood as probing their inner layer where shocks from pulsations and convection cell ejections play an important role. Observations of the $^{29}$SiO(8-7) emission from R Dor and W Hya in a ring of radius 5$<$$R$$<$10 au are illustrated in Figure \ref{fig11}. While both stars display high Doppler velocity wings, partly affected by absorption, the effect is less marked in W Hya than in R Dor, with a $\sigma$ of 4.2 \kms\ instead of 5.2 \kms.

\begin{figure*}
  \centering
  \includegraphics[width=14cm,trim=0.0cm .5cm 0.cm .5cm,clip]{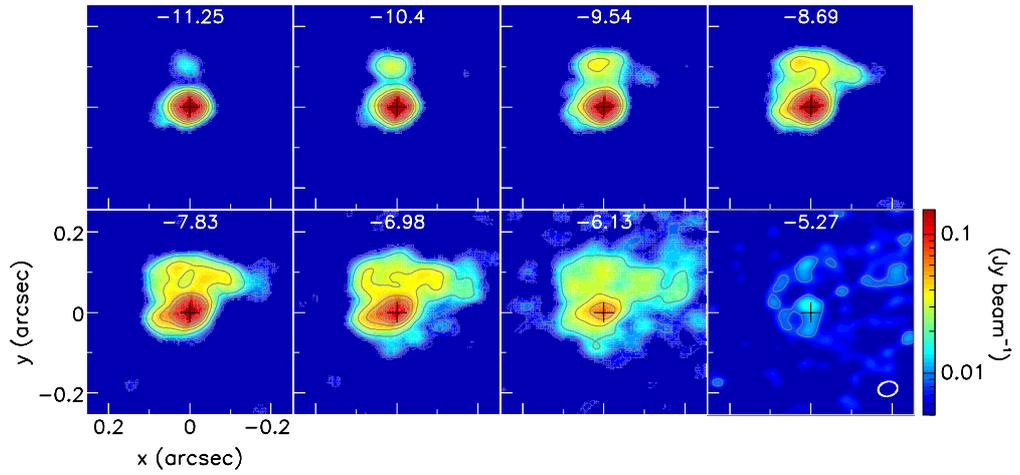}
  \caption{Channel maps of the $^{29}$SiO(8-7) line emission of W Hya in the blue hemisphere showing a blob of enhanced emission. The redmost panel is in the absorption range. Mean Doppler velocities are indicated in each panel. The beam, 52$\times$38 mas$^2$, PA=99\dego\ (robust weighting), is shown in the lower-right corner of the lower-right panel.}
 \label{fig12}
\end{figure*}

\begin{figure*}
  \centering
  \includegraphics[height=5.5cm,trim=0.0cm 1cm 0cm 1.5cm,clip]{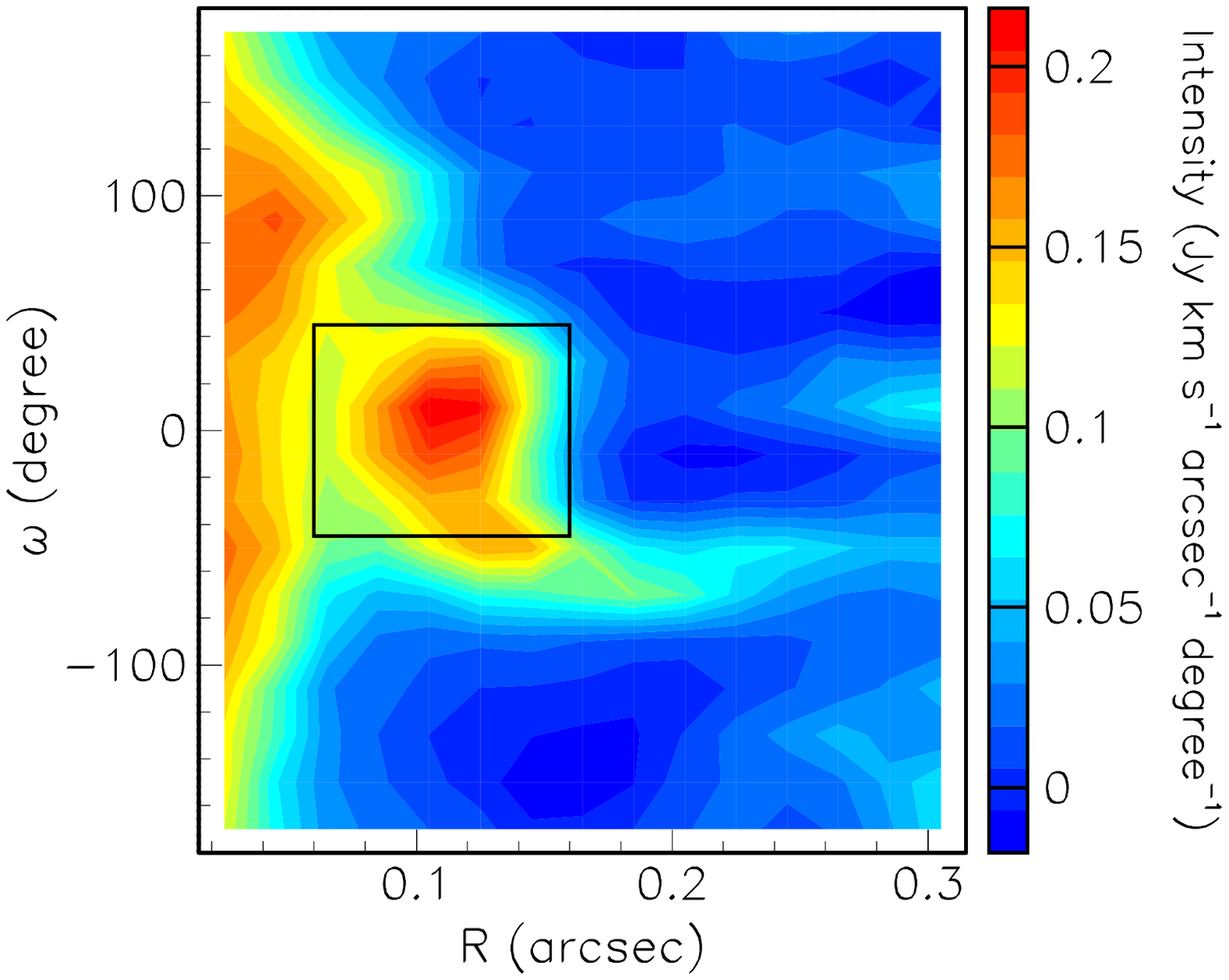}
  \includegraphics[height=5.5cm,trim=0.0cm 1cm 1.cm 1.5cm,clip]{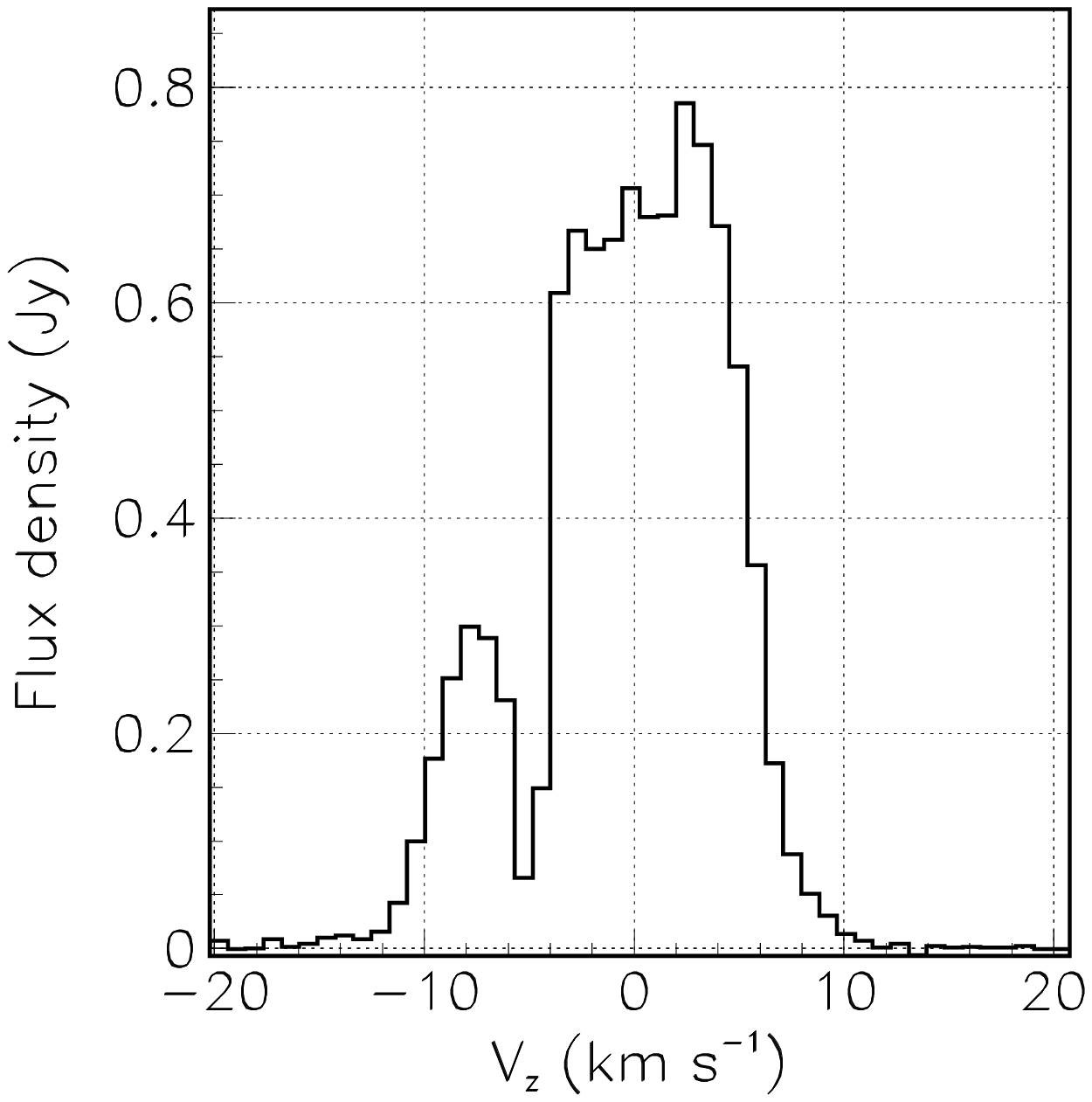}
  \caption{Left: $\omega$ vs $R$ map of the $^{29}$SiO(8-7) line emission of W Hya integrated over the Doppler velocity interval displayed in Figure \ref{fig12}. Right: Doppler velocity spectrum measured in the rectangle shown in the left panel. The beam is the same as in Figure \ref{fig12}.}
 \label{fig13}
\end{figure*}

Finally, we note (Figure \ref{fig12}) the presence of a blob of emission in the blue hemisphere, some 10 au north of the star. It has not been mentioned earlier \citep{Takigawa2017}. Such a blob was first observed in R Dor by \citet{Decin2018} and later shown by \citet{Nhung2021}  to be caused by a stream of gas rather than by a companion as suggested in earlier analyses \citep{Homan2018, Decin2018, Vlemmings2018}. Here, in contrast with R Dor, the distance to the star is independent from Doppler velocity, excluding  a sensible interpretation in terms of a gas flow.  We show in Figure \ref{fig13} the $\omega$ vs $R$ map integrated over $-$11.7$<$$V_z$$<$$-$4.8 \kms\ and the Doppler velocity spectrum integrated over the blob, significantly notched by absorption. The present data show that the blob covers essentially a well-defined compact region of the data cube. However, Figure 13 shows the emission trailing $\sim$60\dego\ west of north toward the region where a blob was observed in the right panel of Figure 8, where imaging is not reliable. Observations using proper $uv$ coverage and possibly including other molecular line emissions would probably help with understanding the nature of the blob.

\section{Summary and conclusions}{\label{sec5}}

The present work, initiated with the intention of exploring the morpho-kinematics of the CSE of W Hya up to distances at arcsec scale using ALMA observations of the $^{29}$SiO(8-7) line, has revealed a major shortcoming preventing such an exploration to be reliably performed. 

 The lack of $uv$ coverage for baselines between 200 m and 400 m has been found to cause 
a strong  distortion of the radial distribution of the detected flux, with a depression at projected distances centred around 0.45 arcsec ($=\lambda/400$ m).  As this can easily be overlooked in other observations using antenna configurations with significant intervals of missing baselines, we devoted the major part of the article to a detailed study of the effect.  Antenna configurations combining an extended array with a compact smaller central array are prone to produce baseline distributions made of two families separated by a gap: short baselines not exceeding the diameter of the central array (here $\sim$200 m) and long baselines between either two antennas of the extended array or an antenna of the extended array and one of the central array (here exceeding $\sim$400 m). Beyond the depression, the observed pattern of emission is governed by the detailed configuration of antennas in the central array; as such, if the $uv$ coverage is inadequate, it may take the form of apparent outflows emitted back-to-back. At variance with real back-to-back outflows that are emitted at opposite Doppler velocities  these artefacts are independent of the frequency interval being considered, providing a useful discrimination against them. In principle, one should be able to cope with this problem by modelling the morpho-kinematics of the CSE and adjusting the associated parameters to best fit the observed visibilities. In practice, however, the complexity of the physics at stake is such that it would be difficult to obtain a convincingly reliable result: the presence of dust formation, of a complex temperature-dependent gas-dust chemistry and of strong absorption prevent such an enterprise to be credibly successful. 

Three earlier publications have been using the same ALMA observations as used in the present article without making explicit reference to the impact of missing baselines. The work of \citet{Vlemmings2017} focuses on the inner part of the CSE, within less than 6 au from the centre of the star, in a region where imaging is perfectly reliable and not at all affected by the missing baselines. Their important conclusions concerning the shock-heated inner atmosphere are therefore fully valid. The work of \citet{Takigawa2017} uses analyses of both AlO and SiO emissions. While the former is confined to the close neighbourhood of the star and is therefore unaffected by the missing baselines, the latter extends further out. Indeed their Figure 2b clearly shows the presence of the depression. However the main conclusion of their work rests on the remark that SiO emission extends much further out than AlO emission. This conclusion not only remains valid but is even strengthened when correcting for the missing flux associated with the missing baselines: their impact is therefore expected to be minimal. Finally, the work of \citet{Danilovich2019}, which studies the abundances of sulfur-bearing molecules CS and SiS, extends well beyond the region where imaging is reliable. It is therefore affected by the depression around 0.45 arcsec while these authors state in their article that ``W Hya was observed on baselines from 17 m to 11 km, giving sensitivity to angular scales up to about 6 arcsec [...] The full width half maximum of the ALMA primary beam is about 15 arcsec in this frequency range and all of the results presented here are from the inner few arcsec where the reduction in sensitivity is negligible.'' It is not up to us to evaluate precisely the impact of the missing baselines on their result. We remark, however, that it is likely not to be very large, being affected only by the distortion of the radial distribution.

The impact of the missing baselines would be considerably more severe on studies that rely on the detailed morphology of the image, not simply on the radial distribution of the intensity. It would result in claiming the presence of back-to-back outflows, which would in fact be pure artefacts.

Finally, while being unable to explore reliably the morpho-kinematics of the CSE up to arcsec scale distances, we have mentioned a few features of lesser importance that were not commented upon in the published literature but are significant enough to deserve a brief presentation. They are presented in the form of a comparison with similar features observed in R Dor and include comments on rotation, absorption and broad line widths. Evidence has been given for the presence of a blob of enhanced emission in the blue-shifted hemisphere, some 10 au north of the star, the present data being however insufficient to propose a reliable interpretation.

\section*{Acknowledgements}
We thank Dr. St\'{e}phane Guilloteau for useful discussions and Dr. Aki Takigawa for clarifications on the maximal recoverable scale used in their article. This paper uses ALMA data ADS/JAO.ALMA\#2015.1.01446.S. ALMA is a partnership of ESO (representing its member states), NSF (USA) and NINS (Japan), together with NRC (Canada), MOST and ASIAA (Taiwan), and KASI (Republic of Korea), in cooperation with the Republic of Chile. The Joint ALMA Observatory is operated by ESO, AUI/NRAO and NAOJ. The data are retrieved from the JVO/NAOJ portal. We are deeply indebted to the ALMA partnership, whose open access policy means invaluable support and encouragement for Vietnamese astrophysics. Financial support from the World Laboratory, the Odon Vallet Foundation and VNSC is gratefully acknowledged. This research is funded by the Vietnam National Foundation for Science and Technology Development (NAFOSTED) under grant number 103.99-2019.368.


\end{document}